\documentclass[aps,prb,twocolumn,floatfix]{revtex4}
\usepackage{graphicx}
\usepackage{epstopdf}
\usepackage{amssymb}
\usepackage{epsf}

\newcommand{\ket}[1]{|#1\rangle}
\newcommand{\braket}[1]{\left\langle#1\right\rangle}

\renewcommand{\phi}{\varphi}
\renewcommand{\epsilon}{\varepsilon}

\renewcommand{\vec}[1]{{\bf #1}}

\newcommand{\eqnref}[1]{Eq.~(\ref{#1})}	
\newcommand{\fref}[1]{Fig.~\ref{#1}}	
\newcommand{\secref}[1]{Section~\ref{#1}}

\newcommand{\beq}{\begin{equation}}
\newcommand{\eeq}{\end{equation}}
\newcommand{\ba}{\begin{array}{ccc}}
\newcommand{\ea}{\end{array}}

\usepackage{amssymb}
\usepackage{amsmath}

\begin{document}
\title{Photo control of transport properties in disorderd wire;\\ 
average conductance, conductance statistics, and time-reversal symmetry}
\author{Takuya Kitagawa$^{1}$, Takashi Oka$^{1,2}$, Eugene Demler$^{1}$} 
\affiliation{$^{1}$Physics Department, Harvard University, Cambridge,
Massachusetts 02138, USA, $^{2}$
Department of Physics, Faculty of Science, University of Tokyo, Tokyo 113-0033, Japan}

\date{\today}
\begin{abstract}
In this paper, we study the full conductance statistics of disordered one dimensional wire under the application of light. 
We develop the transfer matrix method for periodically driven systems to analyze the conductance of 
large system with small frequency of light, where coherent photon absorptions play important role to determine not 
only the average but also the shape of conductance distributions. 
The average conductance under the application of light results from 
the competition between dynamic localization and effective dimension increase, and shows non-monotonic behavior 
as a function of driving amplitude. On the other hand, the shape of conductance distribution 
displays crossover phenomena in the intermediate disorder
strength; the application of light dramatically changes the distribution from log-normal to normal distributions. 
Furthermore, we propose that conductance of disordered systems can be controlled by engineering the shape, frequency and amplitude of light. Change of the shape of driving field controls the time-reversals symmetry and the disordered 
system shows analogous behavior as negative magneto-resistance known in static weak localization. 
A small change of frequency and amplitude of light leads to a large change of conductance, displaying giant-opto response. 
Our work advances the perspective to control the mean as well as the full conductance statistics by coherently driving disordered systems. 
\end{abstract}

\maketitle

\section{Introduction}

The study of disordered systems has been one of the central subjects in physics not only because its direct relevance to 
the material science and technology, but also because it leads to rich 
fundamental physics as first pointed out by Anderson[\onlinecite{Anderson1958}].
While Anderson first considered the problem of non-interacting electrons hopping on a square lattice with 
on-site disorders\cite{Anderson1958}, the study was extended to include the effect of inelastic collisions with phonons\cite{Mott, Ambegaokar1971} and among particles\cite{Basko2006}, to different geometries such as hexagonal lattice in graphene\cite{Wu2007,McCann2006}, and to different physical systems such as
classical light\cite{Wiersma1997,Chabanov2000,Schwartz2007} and neutral cold atoms\cite{Roati2008,Billy2008}. Such extensions not only significantly advanced the understanding of disordered systems and allowed the 
observation of Anderson localization, 
but also they brought many surprises, such as the phenomenon of anti-weak localization 
in graphene\cite{Wu2007,McCann2006} and the presence of insulator-metal transitions in interacting systems\cite{Basko2006}. 
In this paper, we further explore the rich physics of disordered systems by investigating it 
under the application of coherent periodic drive such as light.

Disordered systems in one dimension have been intensively studied in the past, and 
it has been shown through numerical as well as analytical methods that all the states 
in one dimension are localized\cite{Abrahams1979, Beenakker1997}.
 For a finite system, the magnitude of average conductance 
is determined by the ratio of 
localization length $\xi$ and system size $L$; for $\xi/L \ll 1$, the conductance is exponentially suppressed and 
the system is an insulator; on the other hand, for $\xi/L \gg 1$ the system behaves as a conductor.
In this paper, we demonstrate that 
the application of light on such disordered systems can dramatically change the average conductance. 

There are two competing effects that alter the average conductance. One is the effect of dimensional increase. 
In strongly disordered systems, photons can be thought of as providing 
energies to hop from one localized state to another by bridging the energy gaps between the localized states. 
Naively, such photon-assisted hopping decreases the effect of disorders and increases the conductance. 
We show that this photon-assisted hopping provides the effective dimensional increase 
by inducing additional conducting channels called Floquet channels, 
which in return results in the increase of the conductance.
Previously, using such a picture, a pioneering work by  Martinez and Molina[\onlinecite{Martinez2006}] studied the increase of localization length through the low frequency drive. 
The second effect is called dynamic localization\cite{Dunlap1986, Grossmann1991}, which decreases conductance 
when disorders are weak and driving is strong. This effect has been experimentally 
observed in the context of semiconductor superlattices\cite{Keay1995} and neutral cold atoms\cite{Lignier2007}.
The origin of dynamic localization is the decrease of time-averaged hopping amplitudes 
for strong AC driving of the system\cite{Martinez2006,Holthaus1995,Holthaus1996}. 
For generic strength of disorders, these two effects of dimension increase and dynamical localization compete, 
and as a result, the average conductance displays 
non-monotonic behavior as a function of driving amplitude\cite{Martinez2006}.

However, the average conductance is only a partial story; 
in static disordered systems, it is known that higher moments displays rich physics 
such as the universal fluctuations for weakly localized regime\cite{Lee1985, Altshuler1985}, and change of the shape of conductance distributions
for increasing disorder strength. 
In this paper, we study the full conductance distributions to investigate how, for example, the fluctuations of 
conductance changes upon the application of light. 
Recent work by Gopar and Molina\cite{Gopar2010} investigated 
the behavior of conductance distributions for small system size $L=10$, and they 
showed that the distribution can be modified through a sharp cutoff in the high frequency regime. 
Their study employed so-called continued fraction method
to compute the Floquet Green's functions\cite{Martinez2003, Martinez2006, Gopar2010}. This method, being the solution that uses the full matrix for the Hilbert space of $L$ sites, is restricted to small system size.

Light application is expected to yield a large effect
when photon-assisted transport plays an important role, where spatial disorders are avoided through 
the absorptions and emissions of photons. Such photon-assisted transport occurs when 
two sites in the system are coupled resonantly through light. For small system size $L=10$ studied
in the previous work\cite{Gopar2010}, a large driving frequency and amplitude compared to 
the finite size gap $\Delta E \sim 1/L$ were required to satisfy the resonant condition. 
The magnitude of driving amplitude suggested by this work 
is challenging to achieve in experiments.  

In this paper, we demonstrate that 
a large effect of photon-assisted transport 
appears even for small frequency and small amplitude of driving for large system size, 
resulting in the giant opto-response of the conductance in disordered systems. 
In order to study this interesting regime, we develop a transfer matrix method adapted for 
periodically driven systems. This method allows 
us to study $L = 601$ sites when $\sim 10$ photons involve in the transport process. 
Using this formalism, we study the conductance distributions for various intensity of light 
and for various strength of disorders
to give comprehensive picture of conductance behaviors across localized and conducting regimes. 
The analysis of conductance distributions in this regime reveal novel physics that was not observed before; 
a dramatic change of the shape of conductance distributions 
in the intermediate disorder regime, which results in the crossover from log-normal like distribution to normal distributions. 
This result opens the perspective to change the conductance statistics in disordered systems
 through the coherent driving. 
 
 From the point of view of conductance control, active coherent drive through light allows one to effectively change 
 parameters of the system. In this paper, we propose the possibility of changing the symmetry 
 class of driven systems; by  choosing the appropriate shape of the driving, one can preserve or break
 the time reversal symmetry. Time-reversal symmetry plays an important role to determine the conductance
 in the static systems, and we expect similar effects should exist in the driven systems. 
  Here we demonstrate that such control of time-reversal symmetry indeed
 changes the conductance properties in the disordered systems. 
 
 Moreover, in experimental and practical situations, it is interesting to ask if and how much 
 conductance can be controlled by changing the frequency as well as 
 the intensity of light for a sample with a given realization of disorders. 
 Because only slightly different frequency and intensity of light couples very different 
 localized states due to the dense density of states of disordered systems, 
 a small change of driving frequency can induce a large change in the conductance. 
 We will demonstrate, through the formalisms developed in this paper, that the conductance of a sample 
 with a given realization of disorders can be changed by large amount through the small change of 
 frequency and intensity of light. 
 
 While in this paper, we use the language of condensed matter physics, the same consideration
 can be applied to cold atom systems. Recent experiments have studied the transport and localization 
 properties in one and three dimensional disordered cold atom systems\cite{Roati2008,Billy2008, Kondov2011}.
 Moreover, coherent AC drive through shaking is demonstrated\cite{Lignier2007, Chen2011}. 
Due to the versatile controls and probes in cold atom systesms, they provide a promising candidate 
 system to confirm the predictions made in this paper.

This paper is organized as follows. In \secref{sec:summary}, we summarize the main finding of this paper. 
After a brief review of the results of conductance distributions in static disordered systems 
in \secref{sec:review}, we describe the systems 
considered in this paper in \secref{sec:systems}. Here we also discuss the central concepts of this paper, 
dynamic localizations and dimensional increase,  
through the Floquet formalism. In \secref{sec:summary_result},
we give the main results of numerical calculations for the average (\fref{summary_mean}) and full distribution
(\fref{summary_dist}) of conductance, which display different behaviors depending on whether the system is in
strongly localized regime, intermediate regime or conducting regime. 
The detailed formalism of Floquet transfer matrix which allowed the evaluation of conductance distributions 
for large system size is developed in \secref{sec:formalism}. 
In \secref{sec:results}, we supplement the study presented in \fref{summary_mean} and \fref{summary_dist}
through the comparison of average, fluctuation and skewness of conductance distributions for various 
strength of light intensities and disorder. Moreover, we compare these moments of conductance in the 
driven systems with those of static quasi-1D systems for various number of channels and strength of disorder. 
In \secref{sec:TRS} and \secref{sec:giant_response}, we analyze how active control of light leads to the manipulations
of  conductance in disordered systems. As we discussed above, we present in \secref{sec:TRS}
that, by choosing the appropriate shape of the light, one can either break or preserve the time reversal symmetry. 
In \secref{sec:giant_response}, we study how the conductance of a sample with a given realization of disorder 
depends on the frequency and intensity of light. We demonstrate that a large change in the conductance can be 
induced by a small change of either frequency or intensity of light. 


\section{Summary of results} \label{sec:summary}
\subsection{Static disordered systems} \label{sec:review}
The physics of conductance in the disordered systems under the application of light can be best illustrated 
by contrasting with the conductance in the static limit, or the limit in which the intensity of light is set to zero. 
Thus we first summarize the known results in the static systems. 

While the study of disordered systems is challenging in dimensions higher than 
one, elegant analysis through random matrix theory allows the understanding of not only the average value of conducntace, 
but also their full conductance distributions for non-interacting particles in quasi-1D disordered systems
\cite{Beenakker1997}. 
In particular, the solution of Dorokhov-Mello-Pereyra-Kumar (DMPK) equation gives a reliable, analytic 
understanding of conductance distributions in quasi-1D systems\cite{Mello2004}.
Here quasi-1D means the system is one dimension but has width $W$ which is much smaller than the length
of the system $L$, {\it i.e.} $W \ll L$. 
Such analysis shows that a single dimensionless parameter 
$\alpha = \gamma l/L$ determines the conductance distributions in the two
extreme limit of $ \gamma l/L  \ll 1$ and $\gamma l/L \gg 1$, where $l$ is mean-free path, $L$ is the system size
and $\gamma = \beta N + 2 - \beta$ depends on the number of channels/the width of the systems $N$. 
$\beta$ characterizes the symmetry of the systems, and takes 
$\beta =1$ in the presence of time-reversal symmetry 
with no spin-orbit coupling,  and  $\beta=2$ in the absence of time-reversal symmetry. 
When $ \gamma l/L  \ll 1$, quasi-1D system is in localized regime, and the average conductance is exponentially 
suppressed, {\it i.e.} $\braket{ G/G_{0}} \propto \exp\left( -L/(2\gamma l) \right)$ as a function of system size $L$. 
Here $G_{0} = 2e^2/h$ is the natural unit of conductance. In this expression, 
$\xi = \gamma l$ can be seen to correspond to localization length. Moreover, the distribution of the conductance becomes 
log-normal, or, the distribution of log of the conductance $-\ln(G/G_{0})$ becomes Gaussian with mean value
$- \braket{\ln(G/G_{0})} = 2L/\gamma l$ and its fluctuations $\textrm{Var}\left[ \ln( G/G_{0}) \right]  =  4L/\gamma l$ such 
that $P(\ln(G/G_{0})) \propto \exp\left\{ - \left( -\ln(G/G_{0}) - \frac{2L}{\gamma l} \right)^2/ \left( \frac{8L}{\gamma l} \right) \right\}$.
In the other limit $ \gamma l/L  \gg 1$, the system shows Drude-type conducting behaviors with 
average conductance $\braket{G/G_{0}} = N l/L$. The distributions becomes Gaussian and the fluctuation
of $G/G_{0}$ displays the universal features $\textrm{Var}\left[ G/G_{0} \right]  = \frac{2}{15 \beta}$\cite{Beenakker1997}.  
Therefore, conductances of quasi-1D systems increase as one increases the width $N$, and in particular, 
for sufficiently large $N$, there is a crossover from a insulating regime to a conducting regime. 
In analogy with this static results, we show below that the application of light effectively increases the number of channels
$N$ in the system, which lead to similar crossover behaviors. 

We note that the localization length in the strongly disordered regime 
crucially depends on the symmetry of the system such as time-reversal symmetry, 
which determines the value of $\beta$. For $N>1$, the localization length becomes longer when the time-reversal
symmetry is broken.  Moreover, the weak localization in higher dimensions is known to display 
negative magneto-resistance since magnetic field breaks time-reversal symmetry and this destroys the 
constructive interference between time-reversal paths. Thus, time-reversal symmetry controls the 
transport behaviors in disordered systems. We will see in \secref{sec:TRS} that time-reversal symmetry can be controlled 
through the shape of the driving fields in optically driven systems. 


\begin{figure}[t]
\begin{center}
\includegraphics[width = 8.5cm]{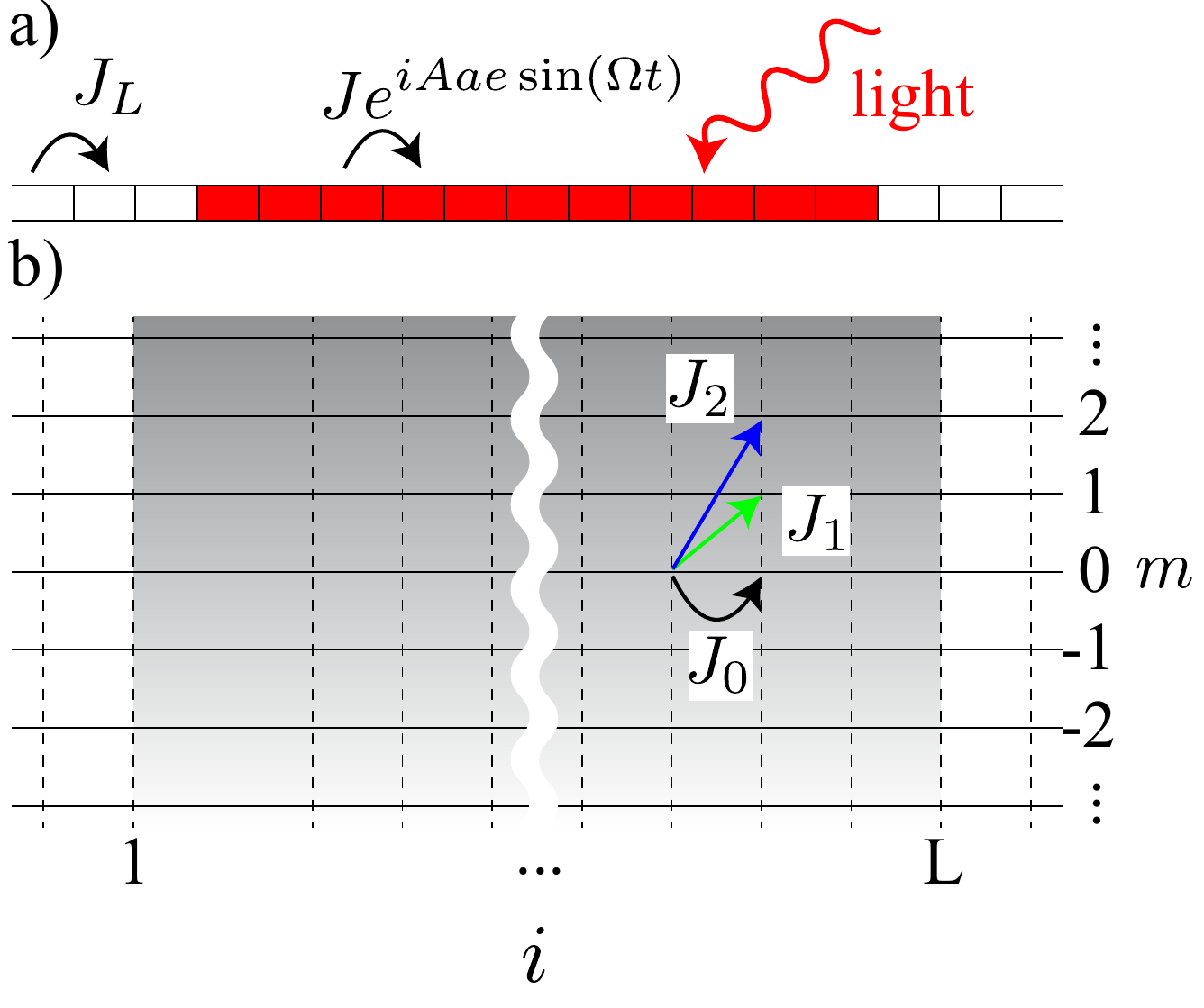}
\caption{a) Illustration of the disordered one dimensional systems under the application of light considered
in this paper. Sites that are driven by light is colored as red, and sites that are not driven by light act 
as leads and colored white. Disorders are also present at sites colored as red, but absent in the white regions corresponding
to the leads. 
 b) 
The one dimensional periodically driven system as illustrated 
in a) can be understood in terms of two dimensional time-independent Hamiltonian of \eqnref{floquetequation} with additional dimension given by energy ladders. A photon can be absorbed or emittted in the system, leading to the hopping 
in the energy ladders. We illustrate in the figure only zero, one and two photon absorption processes, but as 
the intensity of light increases, there are processes which absorbs or emits more photons. The linear potential in 
energy direction coming from the energy difference of different photon number sectors is illustrated by the color 
gradient in the figure. }
\label{scheme}
\end{center}
\end{figure}

\subsection{Systems} \label{sec:systems}
In the following, we consider a tight-binding model for spin-less, non-interacting fermions 
in a one dimensional lattice with on-site disorder potentials $W_{i}$ 
which takes random values between $-W/2$ and $W/2$. The model can be easily extended to include
spin degrees of freedom. We incorporate the effect of the application of light as a phase accumulation
of the hopping. See \fref{scheme}a for illustration. The Hamiltonian is given by 
\begin{equation} \label{hamiltonian}
H(t) = -J \sum_{<ij>} c_{i}^{\dagger} c_{j} e^{i A_{ij}(t)} + \sum_{i} W_{i} c^{\dagger}_{i} c_{i} 
\end{equation}
Here $A_{ij}(t) = e A a \sin(\Omega t) (i-j)$ where $A$ is the amplitude of vector field related to 
the electric field strength as $E = A\Omega$, $a$ is the lattice constant and $e$ is electron charge. 
In the following, we use the units in which $a=1$ and $e=1$. 
Note that for one dimensional lattice with no width, polarization of light does not matter. 
While we assumed that the width of the system is $1$ and thus the number of channels is equal to $1$ 
above, it is straightforward to extend the formalism to the systems with more than one channel. 

Now we consider the Schr\"odinger equation for this time-dependent Hamiltonian. 
We take the Fourier transform of the state which separates the photon absorptions and 
quasi-energy, {\it i.e.} $\ket{\psi(t)} = \sum_{m} \int^{\Omega/2}_{-\Omega/2} dE e^{-it (E+m\Omega)} \ket{\psi_{E}(m)}$.
After the Fourier transform, the Schr\"odinger equation becomes time-independent; 
\begin{equation}
(E+m\Omega) \ket{\psi_{E}(m)} = \sum_{k} H(m-k) \ket{\psi_{E}(k)} \label{floquetequation}
\end{equation}
where $H(m) = \frac{\Omega}{2\pi} \int^{2\pi/\Omega}_{0} dt e^{im\Omega t} H(t)$ is $m$th Floquet component of the Hamiltonian. The state $ \ket{\psi_{E}(m)}$ represents the state with quasi-energy $E$ which has absorbed $m$
number of photons. \eqnref{floquetequation} encapsulates the idea of dimensional increase 
due to the periodic drive, where $H(m)$ acts as a hopping to $m$th neighbor in the energy direction.
We illustrate this picture in \fref{scheme}b. 
In the presence of light with frequency $\Omega$, the energy of the system is not conserved but it can change 
by integer multiples of $\Omega$. This number of absorbed/emitted photons $m$ gives the extra index of states
and the time evolution of states in this one dimensional driven system is effectively the 
evolution of a particle in two dimensional lattice where one dimension comes from spatial dimension 
and another from energy dimension. 
In our system, $H(m) = -J_{m} \sum_{i>j} c_{i}^{\dagger} c_{j} - J_{-m} \sum_{i<j} c_{i}^{\dagger} c_{j}$ where $J_{m} = J B_{m}(eAa)$ with $B_{m}(x)$ is $m$th Bessel function. Here we used the property
of Bessel function that $B_{m}(x) = B_{-m}(-x)$.

One important physics in \eqnref{floquetequation} 
is the presence of $m\Omega$ in the left hand side, where energy difference between
different photon number sectors acts as a linear potential in the energy direction. 
This term leads to the Wannier Stark localization and "Bloch oscillation"
in the energy direction and localizes the particle in a finite number of photon sectors. 
Thus, photon absorptions generically lead to 
increase of conducting channels through increasing photon sectors that are involved in the transport. 
We refer to these channels as "Floquet channels" to distinguish them from channels of static quasi-1D systems. 
The localization length of the Bloch oscillation determines how many Floquet channels are activated for a given frequency and intensity of light. 

A rough estimate of the number of Floquet channels involved in the transport can be given 
in the weakly driven systems with $\mathcal{A} \ll 1$. In the absence of disorder, Hamiltonian becomes diagonal in
quasi-momentum space and takes the form 
\begin{eqnarray*}
H  &=& \sum_{k} - 2 J \cos( k - A\sin(\Omega t)) c^{\dagger}_{k} c_{k} 
\end{eqnarray*} 
For small $\mathcal{A}$, the hopping amplitude $J_{E}$ in the energy direction for a given quasi momentum $k$ is 
approximated by 
$J_{E} = - J\sin(k) B_{1}(\mathcal{A})$. From the study of Bloch oscillations, we know the localization 
length is given by $J_{E}/\Omega \sim J B_{1}(\mathcal{A})/ \Omega$. For the frequency of drive $\Omega =0.01 J$
we consider in this paper, the number of Floquet channels involved in the transport for the intensity of light 
$\mathcal{A} = 0.1 \sim 0.2$ is estimated to be $N_{F} = J B_{1}(\mathcal{A})/ \Omega \approx 5 \sim 10$. 
From the estimate above, we note that 
the localization length increases as the frequency of drive decreases and intensity of light increases. 

Another important effect of the application of light is the behavior of zero-photon absorption/emission
hopping $J_{0}$ in \fref{scheme}b. Since $J_{0} \propto B_{0}(eAa)$, the hopping amplitude decreases 
as the intensity of light is increased. This is the origin of dynamic localization\cite{Dunlap1986, Martinez2006,Holthaus1995, Holthaus1996}. The conductance of the 
system is determined by the competition between the increase of dimension due to the increase in the hopping
amplitude in the energy directions $J_{m}$ with $m \neq 0$ and the dynamic localization due to the decrease 
in the hopping amplitude $J_{0}$. 


The result of such competitions appears in the increase/decrease of conductances in a finite wire. 
Here we consider the attachments of leads to the periodically driven systems and study the DC response current 
for a small chemical potential difference between left and right leads, see \fref{scheme}. We suppose that 
light and disorder exist from site $i=1$ to $i=L$ and refer to this part of the system as "central system." The 
left lead goes from $i=-\infty$ to $i=0$ and the right lead from $i=L+1$ to $i=\infty$. 
Here, we take a simple hopping Hamiltonian in the leads with hopping strength $J_{L}$;
\begin{eqnarray}
H_{L} = - J_{L} \sum_{<ij>}  c_{i}^{\dagger} c_{j} 
\end{eqnarray}
In the following, we assume the same strength of the hopping amplitude $J_{L}$ in both the left and right leads. 
We emphasize that 
attachment of leads is conceptually crucial for determining the occupation numbers of fermions in Floquet states.
In particular, conductances cannot be calculated through Kubo's formula often used in static systems 
unless one makes assumptions about the occupation number of fermions. 

For a periodically driven system, the relation between DC current and the transmission matrix $\hat{t}$ 
is known\cite{Jauho1994, Kohler2005, Moskalets2002, Kitagawa2011}, and given by
\begin{eqnarray}
J &=& J_{\textrm{pump}} + J_{\textrm{res}} \\
J_{\textrm{res}} &=& \sum_{m} |\hat{t}(E_{m}, E_{0})|^2 (\mu_{L} - \mu_{R}) \label{Jres}
\end{eqnarray}
DC current can be separated into two parts. One is a pump current $J_{\textrm{pump}}$ that exists even when 
the chemical potentials of the left lead $\mu_{L}$ and right lead $\mu_{R}$ are the same. When the system 
possesses inversion symmetry, it is necessarily zero. For our disorder system, $J_{\textrm{pump}}$ is generally non-zero
for each realization of disorders, but it is zero on average and we ignore the contribution in this paper. 
Response current $J_{\textrm{res}}$ depends on the transmission amplitude $\hat{t}(E_{m}, E_{0})$ 
of the particle which enters the central system from left lead
with with energy $E_{0} = \mu_{L}$ and leaves to the right lead with energy $E_{m} = \mu_{L}+ m \Omega$. 
This is a natural extension of scattering formalism to the systems that allow absorptions and emissions of 
integer multiples of the photon energy. Here different photon number sectors act as different channels of 
conductions. If we represent $\hat{t}(E_{m}, E_{0})$ as a matrix $\hat{t}$ in terms of the photon numbers, 
then the response conductance $G = J_{\textrm{res}}/ (\mu_{L} - \mu_{R})$ is given by 
$G= G_{0} \left( \hat{t}^{\dagger} \hat{t} \right)_{00}$. This restriction to $00$ component is the physical 
consequence of the fact that the fermions coming from leads are always at the chemical potential $\mu_{L}$. 
 Note the important difference from the expression of conductances for multi-channel static systems 
which sums over all the channels, {\it i.e.} $g = G_{0} \textrm{Tr} \left( \hat{t}^{\dagger} \hat{t} \right)$.
We calculate the transmission amplitudes $\hat{t}(E_{m}, E_{0})$ through the transfer matrix method adapted 
for periodically driven systems. In the following section, we discuss the main results of the analysis and leave
the formalism to \secref{sec:formalism}. 

\subsection{Main results} \label{sec:summary_result}
In this paper, we study the conductance distributions for large system size $L=601$ and small 
driving frequency $\Omega = 0.01 J$. In order to study better the behaviors of conducting regimes, 
we choose the number of (static) channels in the central systems to be $2$ by considering the quasi-1D
systems with width $=2$. This makes the maximum possible conductance to be $2 G_{0}$ in the absence
of disorder. The conductance distributions are calculated for various disorder strength and intensity of light
by repeating the conductance calculations for different realizations of disorders for $10000$ times. 
In the numerical evaluation, it is necessary to terminate the Floquet channels to finite numbers.
We estimated the number of Floquet channels involved in the transport to be $\sim 10$ in \secref{sec:systems} 
and here we retained $20$ Floquet channels to ensure the convergence of conductance.

\subsubsection{Average conductance}

 \begin{figure*}[t]
\begin{center}
\includegraphics[width = 17cm]{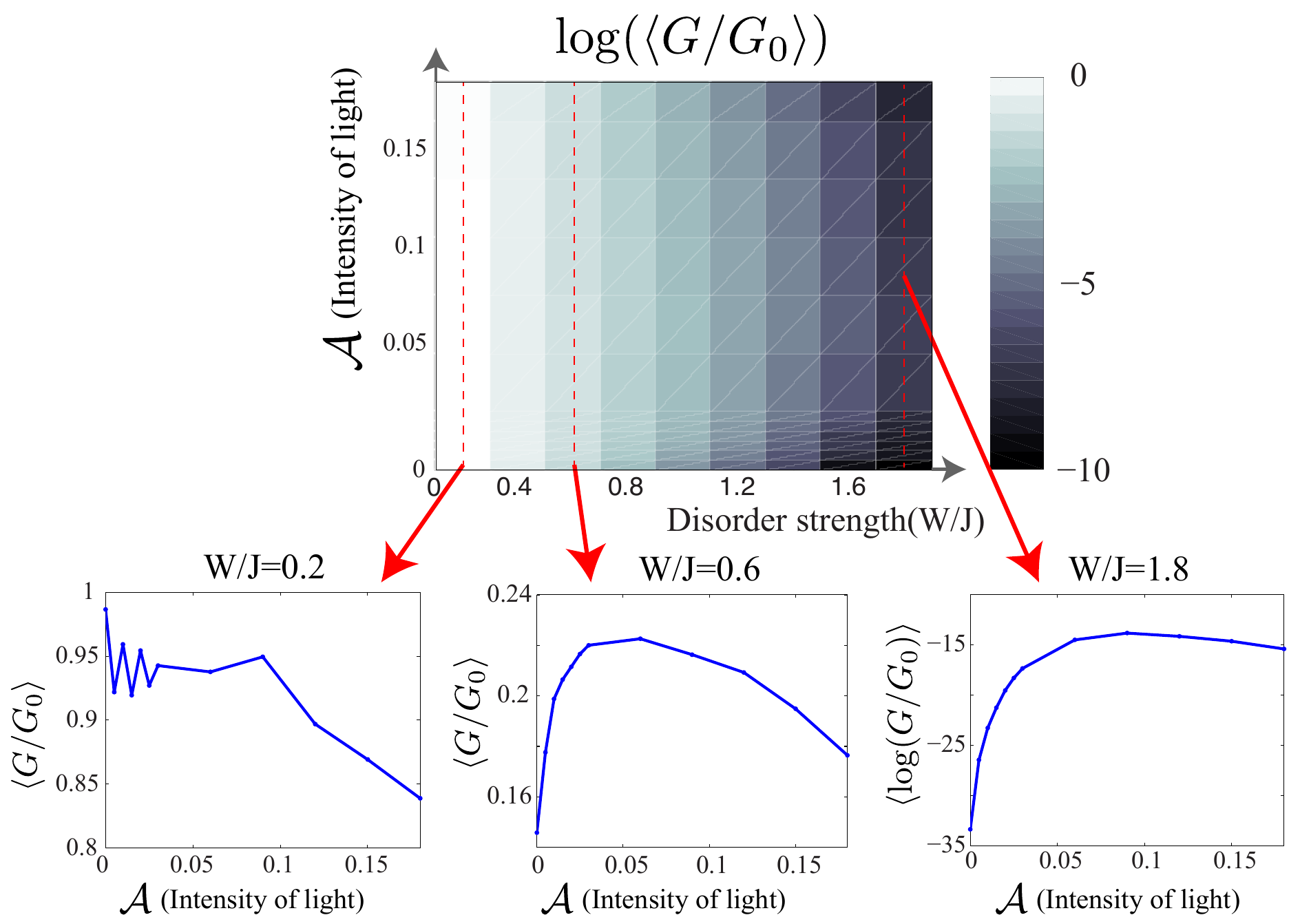}
\caption{(Top) color plot of average conductance $\braket{G/G_{0}}$ for various intensity of light $\mathcal{A} = eAa$
and disorder strength $W$. Here the color is assigned in the log scale. The system size is taken to be $L=601$ with
width (the number of channels) equals $2$, and the driving frequency is $\Omega = 0.01J$.
(Bottom) The line cut of the average conductance
for three different disorder strength, $W=0.2, 0.6, 1.8 J$, are shown at the bottom, corresponding to weak, intermediate and strongly localized regimes. 
For weakly disordered regime $W=0.2 J$ (left), the average conductance decreases for increasing intensity of light $\mathcal{A}$ due to dynamic localization. For intermediate and strongly disordered regimes $W=0.6, 1.8 J$(middle and right), the average conductance increases for small $\mathcal{A}$ due to the increase of the number of Floquet channels. For larger $\mathcal{A}$, 
dynamic localization becomes stronger than the increase of Floquet channels, and the average conductance starts
decreasing. For $W=1.8 J$, we plot the average of logarithmic conductance $\braket{\log(G/G_{0})}$ appropriate
for strongly localized regime.  }
\label{summary_mean}
\end{center}
\end{figure*}

We first analyze the behavior of average conductance for various intensity of light and disorder strength. 
\fref{summary_mean} shows the color plot of the average conductance $\braket{G/G_{0}}$ 
in log scale. The behavior of average conductance as a function of intensity of light, given by 
$\mathcal{A} = eAa$, strongly depends on the disorder strength. 

For weak disorder strength $W=0.2 J$, the localization length is longer than the system size $L$ and 
the conductance is large $\braket{G/G_{0}} \approx 1$. In this regime, application of light leads to
decrease of hopping amplitude $J_{0}$ which is proportional to the zeroth Bessel function 
, {\it i.e.} $J_{0} \propto B_{0}(\mathcal{A})$, leading to dynamic localization 
(see the discussion in \secref{sec:systems}). 
Thus, the conductance tends to decrease as $\mathcal{A}$ is increased as one can see left bottom 
of \fref{summary_mean}. Moreover, the conductance displays oscillations as a function of $\mathcal{A}$. 
Such conductance oscillation has been previously studied in the absence of disorder under AC drive\cite{Martinez2008}, and they have shown that the conductance becomes proportional to 
$B_{0}(\mathcal{A} L)^2$ for large driving frequency. Thus, oscillation period is determined by the system
size $L$. We note that the effect of the dynamic localization is surprisingly large, considering that 
zero-photon hopping is changed only by a small amount at the intensity of light considered here
{\it i.e.} the zeroth Bessel function takes $B_{0}(\mathcal{A} =0.2) = 0.99$. The dependence of conductance
on $B_{0}(\mathcal{A} L)^2$ in the limit of zero disorder and high frequency suggests that 
application of coherent light can decrease the conductance more effectively than 
the naive guess coming from zero-photon hopping strength $J_{0} \propto B_{0}(\mathcal{A} )$.
Detailed analytical study of such giant dynamic localization in the disordered systems is an 
interesting future work. 

On the other hand, 
for intermediate disorder of $W=0.6 J$, the average conductance increases as $\mathcal{A}$ increases for 
small $\mathcal{A}$, see middle bottom of \fref{summary_mean}. 
This increase can be understood from the increase of Floquet channels as we have explained 
in \secref{sec:systems}. However, as the light intensity is further increased, the increase of average conductance 
stops and starts decreasing around $\mathcal{A} \approx 0.05$. Because $\mathcal{A}$ is still small, the number of 
Floquet channels is increasing for increasing $\mathcal{A}$, and thus, such decrease comes from the dynamic localization, where zero-photon absorption/emission hopping amplitude $J_{0}$ is decreased. 
We note that the change of average conductance is relatively small, and the average conductance at $\mathcal{A}=0$
is smaller than that of $\mathcal{A}=0.18$ by about $10 \%$. 
However we will see below that the full conductance distribution shows a dramatic change from 
$\mathcal{A}=0$ to $\mathcal{A}=0.18$. 

For strongly disordered system with $W=1.8 J$, the change of average conductance due to dimensional increase, or,
the increase of the number of channels could be dramatic. In this regime, the conductance $G/G_{0}$ is exponentially
suppressed, so it is more appropriate to consider $\log(G/G_{0})$. In the right bottom of \fref{summary_mean}, 
we have plotted the average of $\log(G/G_{0})$ as a function of $\mathcal{A}$. We see that $\braket{\log(G/G_{0})}$ 
increases from $\approx -33$ at $\mathcal{A} =0 $ to $\approx -15$ at $\mathcal{A} =0.1 $. 
As in the case of intermediate disorder strength, we see the increase of average conductance for small $\mathcal{A}$
and decrease for larger $\mathcal{A}$. Such non-monotonic behaviors of conductance are suggested by previous study of localization length\cite{Martinez2006} where they calculated a similar behavior 
for the localization length as a function of amplitude of driving. 

\subsubsection{Full conductance distribution}

 \begin{figure*}[t]
\begin{center}
\includegraphics[width = 17cm]{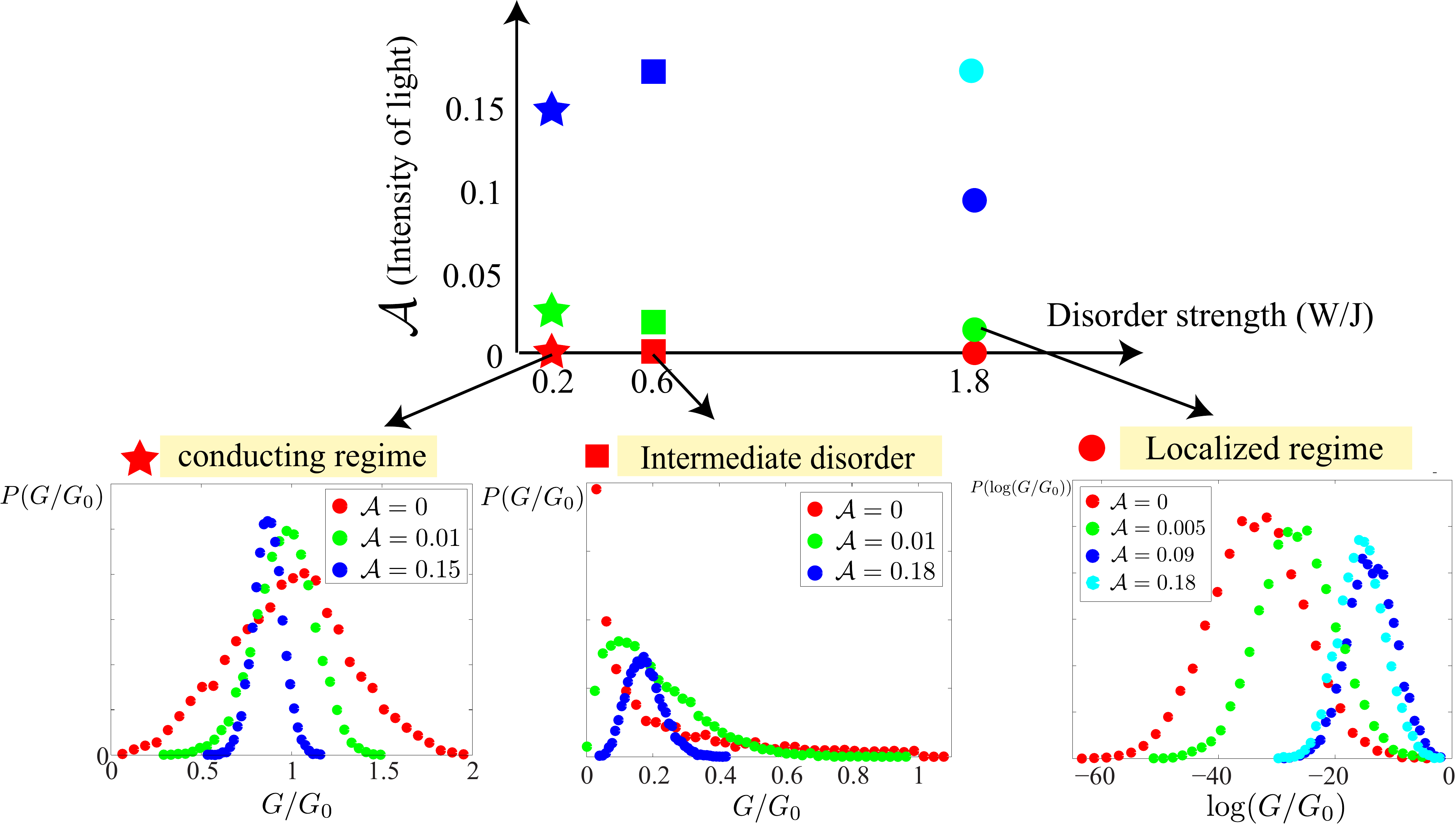}
\caption{Full conductance distributions of disordered 1D systems under the application of light for 
disorder strength $W=0.2 J$(left), $W=0.6 J$(middle), $W=1.8 J$(right) for various intensity of light. 
Each colored marker on the diagram indicates the disorder strength and intensity of light, and 
corresponding conductance distribution is plotted with the same color. For weakly disordered regime (bottom left), 
the conductance distributions take normal distributions, and the shape does not change upon the application of light. 
The average conductance decreases, and moreover, the fluctuation significantly decreases as the intensity of light $\mathcal{A}$ is increased. For strongly disordered regime (bottom right), the distribution of logarithm of conductance is plotted. 
In accord with DMPK equation, the conductance takes log-normal distributions, and this shape does not change 
upon the application of light. As opposed to the weakly disordered regime, the average log-conductance increases
as $\mathcal{A}$ is increased.  The most dramatic change occurs in the intermediate regime, where the conductance
distribution is strongly skewed akin to log-normal distribution in the absence of light. 
As the light is turned on, the distribution immediately changes its shape and approaches normal distribution for larger intensity of light $\mathcal{A}$, displaying crossover behavior.  }
\label{summary_dist}
\end{center}
\end{figure*}

The study of static disordered systems shows rich physics not only in the average conductance 
but also in the full conductance distributions. Motivated by this result, we study the full conductance
distributions under the application of light. 
In \fref{summary_dist}, we have plotted the conductance distributions for various strength of disorder
and intensity of light. 

For weak disorder strength $W=0.2 J$ and zero intensity of light (static limit), 
we see, in agreement with the solution of DMPK equation, that the conductance distribution is
normal distribution (see bottom left of \fref{summary_dist} ). 
As the intensity of light $\mathcal{A}$ is increased from zero,
the average conductance decreases due to dynamic localization, as we have seen above. 
While the shape of distribution stays normal, we see that the width of the distribution, or the fluctuation
of the conductance distributions becomes much smaller. The square root of the variance of $G/G_{0}$ at $\mathcal{A}=0$ 
is $\sqrt{ \textrm{Var}[G/G_{0}]} \approx 0.32$ whereas at $\mathcal{A}=0.15$ 
is $\sqrt{ \textrm{Var}[G/G_{0}]} \approx 0.08$. Considering that the average conductance only changes slightly, which 
goes from $\braket{G/G_{0}} \approx 0.99$ at $\mathcal{A}=0$ to $\braket{G/G_{0}} \approx 0.87$ at $\mathcal{A}=0.15$,
the application of light has more dramatic effects on the fluctuation of the conductance. 

The behavior of average conductance is quite different for weakly disordered regimes and strongly disordered regime
where the former decreases as a function of the intensity of light $\mathcal{A}$, the latter increases.
Nonetheless, strongly disordered regime with $W=1.8 J$ displays a similar behaviors of the decrease of fluctuations as
the intensity of light $\mathcal{A}$ is increased, see the right bottom of \fref{summary_dist}. 
Note that here we plot the distribution of $\log(G/G_{0})$ and not, as in the weakly disordered regime, $G/G_{0}$. 
In accordance with the expectation from DMPK equation, the conductance distribution takes log-normal distributions
in the static limit. The shape of distribution does not change as the intensity of light $\mathcal{A}$ is increased.
While the average of log conductance $\log{G/G_{0}}$ increases, the fluctuation of $\log{G/G_{0}}$ decreases. 

The most dramatic change of conductance distribution occurs in the intermediate disorder regime. 
For zero intensity of light, the distribution of $G/G_{0}$ shows a distribution which has a long tail 
into the large conductance, a signature of log-normal like distribution. As the intensity of light is 
increased, the shape of the distribution immediately changes. It is notable that 
even for a very small intensity of light $\mathcal{A} =0.01$, the part of the distribution with small conductance 
is strongly suppressed and the peak shifts from $G/G_{0}$ to a finite value. As the intensity of light is further increased, 
the distribution approaches normal distribution while the fluctuation also decreases. 

We contrast this behavior of full conductance distribution with 
that for the static quasi-1D disorder systems with varying number 
of channels. 

 \begin{figure}[t]
\begin{center}
\includegraphics[width = 8.5cm]{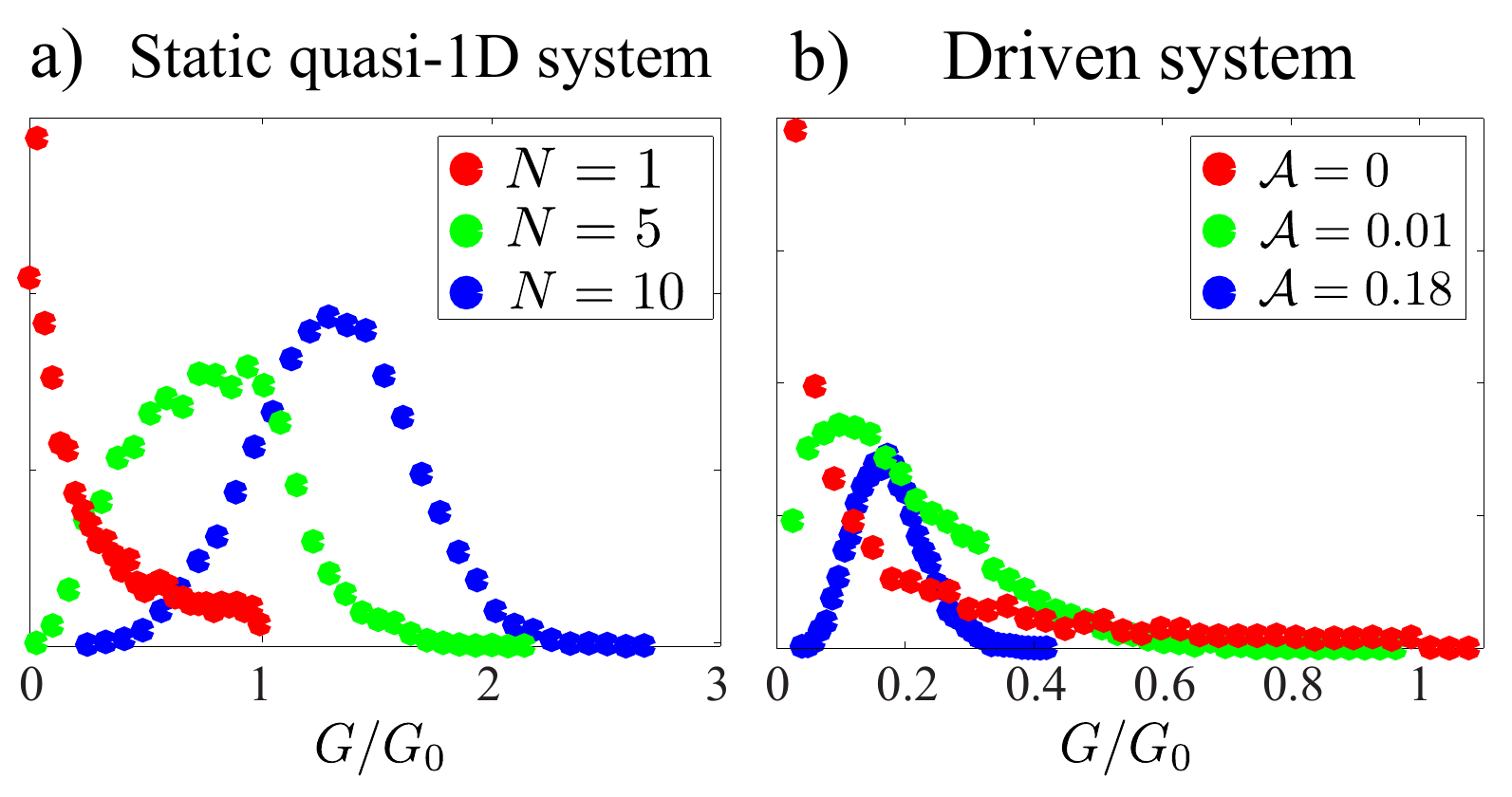}
\caption{a) Conductance distribution of static quasi-1D disordered system for $W=0.4 J$ and various number of 
channels (width). As the number of channels increases, the average conductance increases, and the distribution 
approaches from log-normal to normal distributions. It displays a clear crossover behavior from insulating (log-normal)
to conducting (normal). b) Conductance distribution of driven systems with $W=0.6 J$ for various intensity of light $\mathcal{A}$. While the average conductance does not show a large change, the shape of the distribution changes as $\mathcal{A}$ is increased. This crossover behavior of the shape is quite similar to the crossover observed in the static systems 
for the increasing number of channels.   }
\label{summary_comparison}
\end{center}
\end{figure}

In \fref{summary_comparison} a), we have plotted the full conductance distribution function of 
the static quasi-1D systems with disorder strength $W=0.4 J$ for varying number of channels (width) $N$
of the system. We see that as the number of conducting channels increases, the average conductance 
increases by large amount, which goes from $\braket{G/G_{0}} \approx 0.27$ at $N=1$ to 
 $\braket{G/G_{0}} \approx 1.32$ at $N=10$. 
Such increase of average conductance is accompanied by the dramatic change of the shape, where
the distribution is strongly skewed at $N=1$ but approaches normal distribution at $N=10$. 
Thus as is expected from DMPK equation, the system crossovers from insulating regime for small number 
of channels to conducting regimes for larger number of channels. 

We juxtaposed this distributions for static systems with the conductance distributions under the application
of light for varying intensity of light in \fref{summary_comparison} b). As we have argued in \secref{sec:systems},
the increase of intensity of light $\mathcal{A}$ increases the number of Floquet channels. 
This picture is analogous to the increase of physical channels in the static systems. 

The physical consequence of increase of Floquet channels has similarities as well as differences, compared to
that of increase of physical channels in the static systems. 
First of all, such increase
of Floquet channels leads to only a moderate increase of average conductance.
In fact, as we have seen above, the increase of the average conductance saturates around 
 $\mathcal{A} \approx 0.05$, and it starts decreasing for larger value of $\mathcal{A}$.
 As a result, the average takes $\braket{G/G_{0}} \approx 0.146$ at $\mathcal{A} =0$ and 
 only slightly larger value of $\braket{G/G_{0}} \approx 0.177$ at $\mathcal{A} = 0.18$. 
 On the other hand, the change in the shape of distributions is dramatic, and displays the similar crossover behavior. 
 Thus the increase of Floquet channels in this regime affects the shape of the conductance distribution 
 and induces a crossover from log-normal like distribution to normal distributions. 
 
 While the change of average conductance was moderate for the parameters of our study, 
 it is plausible that for larger number of physical channels, the light application can induce
 larger change. Here we considered $N=2$ case, and this implies the maximum conductance 
 possible is $2$. Increase of the number of channels would increase the maximum conductance,
 and allow larger changes of average conductance. 
 
 Our study of conductance distributions under the application of light has a number of extensions. 
 First of all, the idea of dimensional increase and increase of Floquet channels is general, and applies 
 to any dimension. Static disordered systems are known to be localized in 2D but 
 disordered systems in 3D are believe to have transitions from  localized states to delocalized states, 
 depending on whether the energy of the state is below or above mobility edges. 
 Thus, introduction of coherent light on 2D disordered system can have larger effects on the 
 average conductance, and it is an interesting future work to study this physics.

\section{Floquet transfer matrix method} \label{sec:formalism}
In this section, we give the details of transfer matrix method adapted for periodically driven systems. 
As in \secref{sec:systems}, we choose the gauge in which light application gives rise to 
the hopping of a particle with absorption/emission of energy. In this gauge, when carrying out 
numerical analysis, one often has to terminate the expansion of $e^{i A_{ij}(t)}$
 in a finite harmonics of driving frequency $\Omega$. 
 Equivalently, one can use the gauge in which the light field gives AC 
electric field $E \sum_{i} x_{i} \cos(\Omega t) c^{\dagger}_{i} c_{i} $, which is computationally more convenient for a finite system because there is only the first harmonics of $\Omega$. 
The formalism can be straightforwardly applied to this gauge as well, and we describe the details 
in the Appendix. 

\begin{figure}[t]
\begin{center}
\includegraphics[width = 8.5cm]{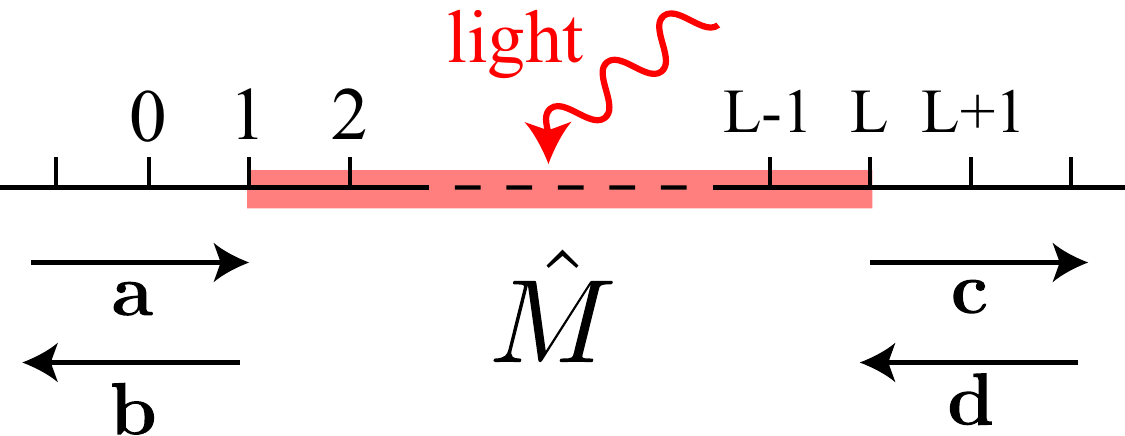}
\caption{ The scattering process for the periodically driven systems. Incoming wave amplitudes and outgoing
wave amplitudes are represented as a vector whose $m$th entry is the amplitude of $m$th Floquet channel 
with energy $E+m\Omega$. Here we apply light to the system with size $L$.  Amplitudes on the left lead is related
to those on the right through the transfer matrix $\hat{M}$, see \eqnref{transfer}. This transfer matrix is directly 
related to the conductance. }
\label{scattering}
\end{center}
\end{figure}

The transmission amplitude $\hat{t}$ relates the amplitudes of incoming modes from the left to 
those of outgoing modes to the right, see \fref{scattering}. In the leads, there is no disorder or 
driving fields, so free propagating mode $e^{\pm ikx}$ is the stationary states with energy $E= -2J_{L} \cos(k)$. 
We represent the wavefunction on the left leads as 
$ \psi_{L}(x)=\frac{1}{\sqrt{2 \sin(\hat{K})}} (e^{-i\hat{K}x} \vec{a} + e^{i\hat{K}x} \vec{b} )$ 
and similarly the wavefunction on the right leads as 
$  \psi_{R}(x)= \frac{1}{\sqrt{2 \sin(\hat{K})}} ( e^{-i\hat{K}x}\vec{c}+e^{i\hat{K}x} \vec{d} )$.
Here the components of the amplitude vector $\vec{a}$ correspond to Floquet channels, where $m$th entry 
is the amplitude of states with energy $E + m\Omega$. The matrix $\hat{K}$ is the the diagonal matrix
whose diagonal vector gives the wavenumber $k_{n} = \cos^{-1}( (E + n\Omega)/(-2J_{L}))$ for each 
Floquet state. The normalization of the state $\sqrt{2 \sin(\hat{K})}$ ensures the unitarity 
of scattering matrix, {\it i.e.} $\hat{S} \hat{S}^{\dagger} = \hat{1}$,  defined below. 
Because the leads have 
finite bandwidth, the transmission of particles has a hard cutoff at a finite Floquet level $|n| \approx 2J_{L}/\Omega$. 
Throughout the paper, we take $J_{L}$ to be much larger than $J$ such that all the Floquet states occupied in the 
central system can be smoothly transmitted to the leads. Even when $J_{L}$ is not much larger than $J$, this 
gives a good approximation if the transmission amplitudes $\hat{t}(E_{m}, E)$ for $E_{m}> 2J_{L}$ are small. 

Then the amplitudes $\vec{a}, \vec{b}, \vec{c}, \vec{d}$ are related through  scattering matrix 
$\hat{S} = \left( \begin{array}{cc} \hat{r} & \hat{t}' \\ \hat{t} & \hat{r}'  \end{array} \right)$ as 
\begin{equation}
\left( \begin{array}{c} \vec{b} \\ \vec{c} \end{array} \right) = 
\left( \begin{array}{cc} \hat{r} & \hat{t}' \\ \hat{t} & \hat{r}'  \end{array} \right)
\left( \begin{array}{c} \vec{a} \\ \vec{d} \end{array} \right) 
\end{equation}
Computationally convenient, closely connected matrix is the transfer matrix $\hat{M}$ which 
relates the amplitudes on the left to the amplitudes on the right; 
\begin{equation} \label{transfer}
\left( \begin{array}{c} \vec{d} \\ \vec{c} \end{array} \right) = 
\hat{M}
\left( \begin{array}{c} \vec{b} \\ \vec{a} \end{array} \right) 
\end{equation}
It can be easily checked that $\hat{M}$ can be expressed through the elements of 
scattering matrix $\hat{S}$ as 
\begin{eqnarray}
\hat{M} = \left( \begin{array}{cc} \hat{t}^{-1} & - \hat{t}^{-1} \hat{r}'  
\\  \hat{r} \hat{t}^{-1} & \hat{t}' - \hat{r} \hat{t}^{-1} \hat{r}'  \end{array} \right)
\end{eqnarray}
In the following, we relate the Hamiltonian to $\hat{M}$, and obtain the expression of $\hat{t}$ 
through the relation above. 

It is straightforward to write down the local transfer matrix $\hat{M}_{i}$ which "transfer" 
the amplitudes of wavefunction from site $i$ to site $i+1$. At each site $i$, the wavefunction 
has components with different Floquet levels. The amplitude associated with each level 
is just the projection of $\ket{\psi_{E}(m)}$ on site $i$, {\it i.e.} $s_{m i} = \braket{i | \psi_{E}(m)}$. 
We represent the amplitude of wavefunction at site $i$ as a vector $\vec{s}_{i}$ whose $m$th
component is $s_{m i}$. 
If we write the Schr\"odinger equation 
\eqnref{floquetequation} in terms of the amplitudes of wavefunction at each site $\vec{s}_{i}$,
then 
 \begin{eqnarray*} 
  (E + \hat{\Omega} - W_{i}) \vec{s}_{i} &=& - \hat{J}^{T} \vec{s}_{i-1} -\hat{J}  \vec{s}_{i+1} \\
   \hat{\Omega} &=& \left(
\begin{array}{ccccc}  \ddots  & &   &  &   \\ &1 \Omega& && 
\\    &  & 0 \Omega & &
 \\  & & & -1 \Omega &   \\  & & &   & \ddots 
\end{array} \right)  \\
\hat{J} &=&  \left(
\begin{array}{ccccc}  \ddots  & &   &  &   \\ &J_{0} &J_{1}  &J_{2} & 
\\    & J_{-1}  & J_{0} & J_{1} &
 \\  &  J_{-2}& J_{-1} & J_{0} &   \\  & & &   & \ddots 
\end{array} \right)
  \end{eqnarray*}
Here, $J_{m} = J B_{m}(eAa)$ is a hopping with $m$ photon absorptions 
where $B_{m}(x)$ is $m$th Bessel function. 
Formally, this equation can be written in terms of the transfer matrix $\hat{M}_{i}$ as 
 \begin{eqnarray*} 
  \left( \begin{array}{c} \vec{s}_{i+1} \\ \vec{s}_{i} \end{array} \right) &=& 
\hat{M}_{i} \left( \begin{array}{c} \vec{s}_{i}  \\ \vec{s}_{i-1}  \end{array} \right) \\
\hat{M}_{i} &=&
\left( \begin{array}{cc} -\hat{J}^{-1} (E + \hat{\Omega} - W_{i}) & -\hat{J}^{-1} \hat{J}^{T} 
\\ 1 & 0  \end{array} \right)
\end{eqnarray*}
  
  If we suppose that the size of the central system to be $L$ 
  where disorder and driving fields exist only from site $1$ to site $L$, 
  then the transfer matrix from sites ($0$, $1$) to sites ($L$, $L+1$)
  is given by $\hat{M}_{1L} = \hat{M}_{L} \prod_{i=2}^{L-1} \hat{M}_{i} \hat{M}_{1}$. 
  Here the transfer matrix that connects the central system and left and right leads, 
  $\hat{M}_{1}$ and $\hat{M}_{L}$ respectively, are given by 
   \begin{eqnarray*} 
\hat{M}_{1} &=&
\left( \begin{array}{cc} -\hat{J}^{-1} (E + \hat{\Omega} - W_{i}) & -\hat{J}^{-1} J_{L}
\\ 1 & 0  \end{array} \right) \\
\hat{M}_{L} &=&
\left( \begin{array}{cc} - J_{L}^{-1} (E + \hat{\Omega} - W_{i}) & - J_{L}^{-1} \hat{J}^{T} 
\\ 1 & 0  \end{array} \right)
\end{eqnarray*}
  Note that whereas $\hat{M}_{1L}$ connects the amplitudes $\vec{s}_{i}$ 
  at sites i=$1$, $0$ to sites i=$L+1$, $L$, the transfer matrix  $\hat{M}$ defined above connects 
 amplitudes $\vec{a}, \vec{b}, \vec{c}, \vec{d}$ of the free propagating mode in the leads from the left lead to the right lead. 
 These two different amplitudes are related through the equations 
 \begin{eqnarray}
 \left( \begin{array}{c} \vec{s}_{1} \\ \vec{s}_{0} \end{array} \right) &=& \hat{Q}
\left( \begin{array}{c} \vec{b} \\ \vec{a} \end{array} \right) \\
 \left( \begin{array}{c} \vec{s}_{L+1} \\ \vec{s}_{L} \end{array} \right) &=& \hat{Q}
\left( \begin{array}{c}e^{i\hat{K} L} \vec{d} \\ e^{-i\hat{K} L}\vec{c} \end{array} \right) \\
\hat{Q}&=& \frac{1}{ \sqrt{2 \sin(\hat{K})}} \left( \begin{array}{cc} e^{i\hat{K}} & e^{-i\hat{K}} \\ \hat{1} & \hat{1}  \end{array} \right)
 \end{eqnarray}
 Thus, the transfer matrix $\hat{M}$ is obtained through $\hat{M}_{1L}$ as 
 \begin{equation}
 \hat{M} =\left( \begin{array}{cc} e^{-i\hat{K}L} & 0 \\ 0 & e^{i\hat{K}L}  \end{array} \right)
  \hat{Q}^{-1} \hat{M}_{1L} \hat{Q}
  \end{equation}
  The unitary transformation $\left( \begin{array}{cc} e^{-i\hat{K}L} & 0 \\ 0 & e^{i\hat{K}L}  \end{array} \right)$
  has no effect on the physical transport properties, so in the following, we study 
  the equivalent transfer matrix $ \hat{M} = \hat{Q}^{-1} \hat{M}_{1L} \hat{Q}$. 
  
  In principle, it is now straightforward to calculate $\hat{M}$ to obtain the conductance of the driven system
 $G= G_{0} \left( \hat{t}^{\dagger} \hat{t} \right)_{00}$. Numerically, this procedure faces difficulty because
 calculation of $ \hat{M}$ gives the inverse of $ \hat{t}$ and one has to take the inversion. Since states of disordered
 one dimensional system are localized, many transmission amplitudes are exponentially small as a function of system size.
In order to have a reasonable precision in the computations of transmission amplitudes, 
it is necessary to keep the small numbers in 
 the inverse matrix $ \hat{t}^{-1}$ which gives a large contribution to the transmission amplitude 
 $ \left( \hat{t}^{\dagger} \hat{t} \right)_{00}$. However, since the transmission amplitudes are very small 
 in the disordered regimes, it is generically necessary to maintain accuracies that retain exponentially small
 numbers. This faces a difficulty as one goes into deeper localized regimes. 
  
 This problem arises since  $\hat{M}$ is the multiplication of many matrices with disorder at each site, which
 creates eigenvalues that are separated by exponentially large numbers. 
 An ingenious method to avoid the problem is to take the inverse before the multiplication\cite{pendry1992, Markos2006}. 
 We aim to obtain the inverse of the matrix 
 $ \hat{t}^{-1} = \left( \begin{array}{cc} 1 & 0 \end{array} \right)  \hat{Q}^{-1} \hat{M}_{1L} \hat{Q} \left( \begin{array}{c} 1 \\ 0 \end{array} \right)$. To this end, we first start from 
 $\left( \begin{array}{c} u_{1} \\ d_{1} \end{array} \right) =  \hat{Q} \left( \begin{array}{c} 1 \\ 0 \end{array} \right)$. 
 Now we rewrite  $\left( \begin{array}{c} u_{1} \\ d_{1} \end{array} \right) =  \left( \begin{array}{c} u_{1} d_{1}^{-1} \\ 1 \end{array} \right) d_{1}$. 
 In the next step, we multiply the transfer matrix $\hat{M}_{1}$ to obtain
 $\left( \begin{array}{c} u_{2} \\ d_{2} \end{array} \right)= \hat{M}_{1} \left( \begin{array}{c} u_{1} d_{1}^{-1} \\ 1 \end{array} \right) $ We repeat this procedure to obtain the equality 
 \begin{equation}
 \left( \begin{array}{cc} 1 & 0 \end{array} \right)  \hat{Q}^{-1} \hat{M}_{1L} \hat{Q} \left( \begin{array}{c} 1 \\ 0 \end{array} \right)
 = u_{L+1} d_{L} \cdots d_{1}
 \end{equation}
 The transmission amplitude is then given by 
  \begin{equation}
 \hat{t} = d_{1}^{-1} d_{2}^{-1} \cdots d_{L}^{-1} u_{L+1}^{-1}
 \end{equation}
 
 From the expression of \eqnref{Jres}, the conductance and Floquet transmission amplitude 
$\hat{t}^{\dagger}$ are related by the equation 
\begin{equation}
G= G_{0} \left( \hat{t}^{\dagger} \hat{t} \right)_{00} \label{conductance}
\end{equation}
where $G_{0}$ is the unit of conductance $e^2/h$ in the case of spinless electron in our study. 
As we have noted in the summary section, it is crucial that the the equation \eqnref{conductance}
is only $00$ Floquet component of the Floquet transmission probability $\hat{t}^{\dagger} \hat{t} $.
This is in stark contrast with the static, multichannel case, where the conductance is given by
the trace of multi-channel transmission probability because incoming electrons can be in any of 
the channels present. In the Floquet channel case, incoming electrons
are all sitting near the Fermi surface, which restrict the conductance to be $00$ Floquet component of 
$\hat{t}^{\dagger} \hat{t}$. 

 \section{Result} \label{sec:results}
 
  \begin{figure*}[t]
\begin{center}
\includegraphics[width = 17cm]{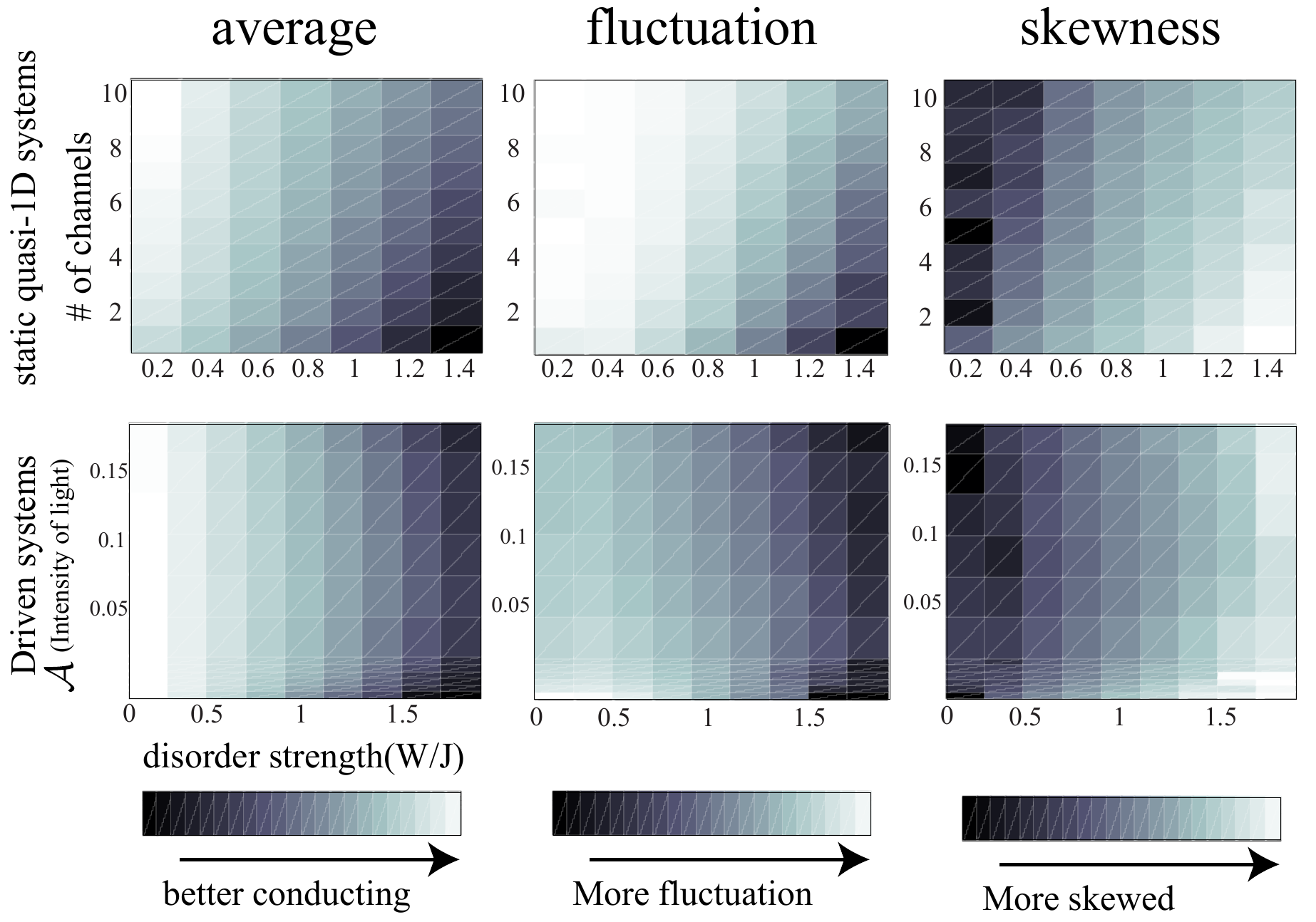}
\caption{Color plot of average, fluctuations and skewness (see \eqnref{skewness})) for static quasi-1D systems
and driven systems. Here, the color is assigned in the log scale.  For static quasi-1D systems (top), the behavior of 
all moments is governed by the crossover from conducting regime (normal distribution with large average conductance)
to insulating regime (log-normal distribution with small average conductance). Such crossover region in the diagonal 
line is apparent. On the other hand, driven systems display richer structure (bottom). 
The average conductance shows little change in the 
weakly disorder regimes as $\mathcal{A}$ is increased, it increases for small $\mathcal{A}$ and decreases for larger 
$\mathcal{A}$ in the strongly disordered regime. The fluctuation is strongly suppressed as $\mathcal{A}$ is increased 
from zero for weakly and intermediate disordered regimes. Skewness displays a similar crossover behavior as 
that of static quasi-1D systems, as is apparent also in \fref{summary_dist}. 
}
\label{summary_colorplot}
\end{center}
\end{figure*}
 
 \subsection{Analysis of moments} 
 In \secref{sec:summary_result}, the results indicate that the application of light not only changes the 
 average conductance, but also significantly changes the shapes of conductance distributions. 
 One way to characterize the shapes of distributions is through their moments. Each moment 
 captures a particular aspect of the distributions, and we will analyze each moment in this section. 
 
 The first moment $\braket{G/G_{0}}$ gives the average conductance, which we have argued, 
 results from the competition between dimensional increase and dynamic localization.  
 Because of this competition, behaviors upon the increase of intensity of light are different from 
 the increase of physical channels in the static systems. In \fref{summary_colorplot} a) and b), we plot
 the color plot of average conductance in the log scale,  \fref{summary_colorplot} a) is the average conductance
 for static quasi-1D systems for various disorder strength and the number of physical channels. 
 As is expected, the average conductance increases as the number of channel increases and disorder strength
 decreases. There is a clear crossover from insulating to conducting behaviors in the diagonal of the color plot.
 This behavior should be contrasted with \fref{summary_colorplot} b), where the average conductance is plotted
 for various intensity of light $\mathcal{A}$ and disorder strength. In contrast with the static counterpart, 
 the average conductance does not show large increase, except for the strongly disordered regimes. 
 For large intensity of light, the decrease of conductance due to dynamic localization is recognizable. 
 
 Second moments gives the fluctuation, given by $\braket{(G/G_{0} - \braket{ G/G_{0}})^2}$. 
 As is evident from the distributions plotted in \fref{summary_dist}, 
 the application of light significantly reduces the fluctuation in the weakly disordered regimes. 
 Motivated by this observation, we plot the fluctuations as a color plot in \fref{summary_colorplot} c) and d)
 where color is assigned in the log scale. For comparison, 
 we plot in \fref{summary_colorplot} c) the strength of fluctuations of static 
 quasi-1D systems for various  disorder strength and the number of channels. Fluctuation generally 
 increases as the number of channels increases because the average conductance also increases. 
 In \fref{summary_colorplot} d), we plot the strength of fluctuations of systems under the application of light 
 for various intensity of light and disorder strength. 
In contrast to static systems, while the average conductance increases as the intensity of light is increased 
(see \fref{summary_dist}), the fluctuation is decreased as the light intensity is increased from zero. 

Third moment leads to the skewness, which is the measure of how symmetric the distribution is. 
Concretely, skewness is defined as 
\begin{eqnarray}
\kappa = \frac{ \braket{ (G/G_{0} - \braket{ G/G_{0}})^3 }}{ \braket{(G/G_{0} - \braket{ G/G_{0}})^2}^{3/2}} \label{skewness}
\end{eqnarray}
 In \fref{summary_colorplot} e) f), we plot
the skewness of the conductance in the log scale color plot for static and driven systems, respectively. 
Skewness can take either positive or negative value, and negative skewness means that the distribution
has long tails on the left, and positive skewness means the distribution has long tails on the right. 
Here we plot the absolute value of the skewness in the log scale. 

In the static quasi-1D systems, skewness increases as the disorder strength is increased and 
it decreases as the number of channels is increased, see \fref{summary_colorplot} e). 
Both effects can be understood as the result of 
transition from normal to log-normal distributions as the system goes from conducting regime to localized regime. 
 The skewness of driven systems has the tendency to decrease as the intensity of light is increased and 
 the disorder strength is decreased. The dependence on the disorder strength can be understood 
 in a similar fashion as the static systems as the crossover between insulating regime and conducting regime. 
 On the other hand, the decrease of skewness on the increase of intensity has the larger effects 
 than what is implied by the change of average conductance (also compare with \fref{summary_dist}). 
 
 For both the fluctuations and skewness, the behaviors of moments for the static quasi-1D systems 
 are tied with the average conductance; the crossover behaviors from conducting regimes (left top) to the
 insulating regimes (right bottom) are apparent in any of three color plots of average, fluctuation and skewness. 
On the other hand, one of the main observations of our study for disordered systems under the application of light is 
the behaviors of average, fluctuation and skewness are not necessarily correlated; fluctuations and skewness 
are strongly suppressed by the application of light when the average conductance decreases or barely changes
in the weak and intermediate disorder strength. 

\subsection{Control of time-reversal symmetry through the driving shape} \label{sec:TRS}
In the previous sections, we consider the application of a simple harmonic drive. 
Such drive which possesses the time $\tau$ such that theHamiltonian is symmetric around that 
point $H(\tau-t)^{*} = H(\tau+t)$ for all $t$ can be shown to be "time-reversal symmetric" 
when the dynamics is "averaged" over one period of time\cite{Kitagawa2010}.  Here $H^{*}$ is the complex conjugation
of the Hamiltonian and we assumed that the electrons are spin-less. 
Such time-reversal symmetry plays an important role in the static disordered systems, and the presence and absence
of time reversal symmetry changes the average as well as fluctuations of conductance, 
as we have described in \secref{sec:review}. 
In this section, we discuss the possibility of changing the symmetry class of the driven disordered systems 
by choosing the shape of the driving. 

\begin{figure}[t]
\begin{center}
\includegraphics[width = 8.5cm]{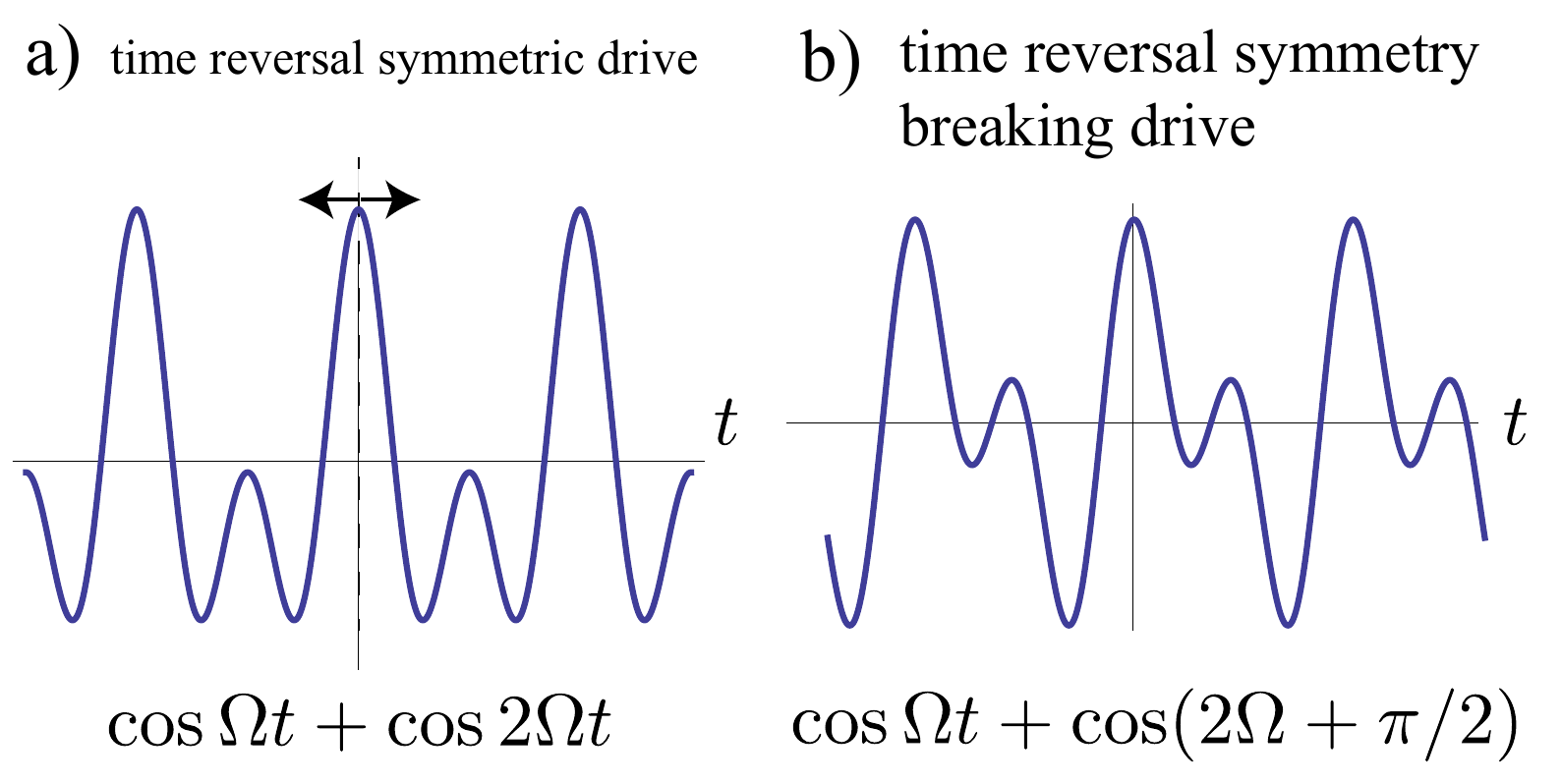}
\caption{The shape of driving field can modify the symmetry class of the driven systems. 
Here, the driving of the form $\cos \Omega t +  \cos \left( 2 \Omega t  + \phi \right)$ is considered.
a) $\phi=0$ leads to symmetric shape of the drive, which implies the presence of time-reversal symmetry. 
b) when $\phi \neq 0$ or $\pi$, the drive breaks time-reversal symmetry. Here we plot the shape of the drive
for $\phi = \pi/2$.   }
\label{drive_comparison}
\end{center}
\end{figure}

It is possible to break the time reversal symmetry of the Hamiltonian by changing the shape of the driving, 
for example, by driving the systems with frequency $\Omega$ and $2 \Omega$ with some phase difference $\phi$. 
When $\phi=0$, the composite drive again possesses a symmetric time $\tau$, but for generic values of $\phi$, 
there is no symmetric time $\tau$, see \fref{drive_comparison}. Intuitively, this is an indication of 
the lack of extra symmetry in the driving.

From random matrix point of view, the difference between the two drives can be seen as 
the difference in the symmetry class of Floquet Hamiltonian. First we rewrite \eqnref{floquetequation}
as a matrix equation for each Floquet level as 

\begin{widetext}
\begin{equation}
E  \left(
\begin{array}{c} \vdots  \\  \ket{\psi_{E}(1)} \\ \ket{\psi_{E}(0)} \\ \ket{\psi_{E}(-1)} \\ \vdots  
\end{array} \right) = 
\left\{ -  \left(
\begin{array}{ccccc}  \ddots  & &   &  &   \\ &1 \Omega& && 
\\    &  & 0 \Omega & &
 \\  & & & -1 \Omega &   \\  & & &   & \ddots 
\end{array} \right) + \left(
\begin{array}{ccccc}  \ddots & \vdots  &  \vdots  & \vdots   &   \\ \cdots & \hat{H}_{0} & \hat{H}_{1} && 
\\    &  \hat{H}_{-1} & \hat{H}_{0} & \hat{H}_{1} &
 \\  & & \hat{H}_{-1} & \hat{H}_{0} &  \cdots \\  & \vdots  &  \vdots  & \vdots   &  \ddots
\end{array} \right) \right\} 
 \left(
\begin{array}{c} \vdots  \\  \ket{\psi_{E}(1)} \\ \ket{\psi_{E}(0)} \\ \ket{\psi_{E}(-1)} \\ \vdots  
\end{array} \right) \label{floquethamiltonian} 
\end{equation} 
\end{widetext}
This equation can be thought of as the eigenvalue problem for Floquet states and 
we refer to the matrix appearing in the right hand side of \eqnref{floquethamiltonian} which
multiplies Floquet states as Floquet Hamiltonian $\vec{H}_{F}$. 
We consider the time reversal symmetry of the driven system as the straightforward extensions of 
time reversal symmetry of static systems applied to $\vec{H}_{F}$; time-reversal symmetry is the existence of anti-unitary 
operator $\mathcal{T}$ such that $\mathcal{T} \vec{H}_{F} \mathcal{T}^{-1} =  \vec{H}_{F}$. 

If we represent the driving by light as the time-varying electric potential as in Appendix, then 
it is clear that the simple drive $E \cos(\Omega t)$ leads to $\vec{H}_{F}$ that is real, and therefore, 
the system is time reversal symmetric with $\mathcal{T} = \vec{1} $.
Now we add the second driving with doubled frequency $E \cos(2\Omega t + \phi)$. When 
$\phi=0$,  $\vec{H}_{F}$ is again real. On the other hand, for $\phi \neq 0$ or $ \pi$, $\vec{H}_{F}$ becomes
complex due to the relative phase between $\Omega$ drive and $2\Omega$ drive, and the time reversal
symmetry is broken. Because the symmetry class of Floquet Hamiltonian is different for these drives, 
the conductance property of these two drives are expected to be different as in the case of static systems. 

\begin{figure}[t]
\begin{center}
\includegraphics[width = 8.5cm]{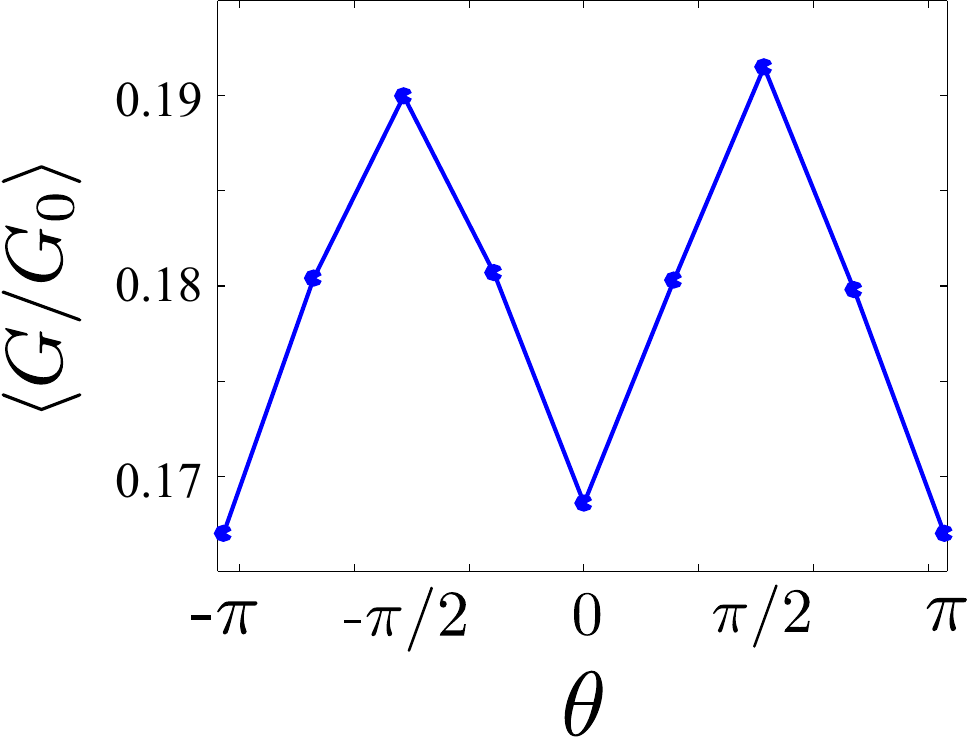}
\caption{ The average conductance of disordered systems in the presence of two driving 
$E \left\{ \cos(\Omega t)  + \cos(2\Omega t + \phi) \right\}$ for various values of $\theta$. 
The dot is the numerical result and the line is the guide to the eye. 
Here, $L=601$, $W=0.6 J$, $\Omega=0.01 J$ and intensity of light $\mathcal{A} =0.3$. 
Similar to the negative magnetoresistance of weak localization, the driven system also 
displays the increase of conductance under the breaking of time-reversal symmetry. }
\label{theta_variation}
\end{center}
\end{figure}
In \fref{theta_variation}, we observe the variation of average conductance for 
disordered systems in the presence of two driving frequencies $\Omega$ and $2\Omega$
with various phase difference between the two.
In terms of electric potential, the form of the driving is 
$E \left\{ \cos \Omega t +  \cos \left( 2 \Omega t  + \phi \right) \right\}$. 
Here we took the intensity of light $\mathcal{A} =0.3$, $\Omega =0.01 J$, $W =0.6 J$, $L=601$.
In the calculation, we retained the number of Floquet states $\pm 50$ to ensure the 
convergence of the conductance. 
$\theta$ can be thought of as "magnetic field" that breaks the time-reversal symmetry. 
Similar to the weak localization in static disordered systems, the driven systems displays
negative "magnetoresistance," in which the breaking of time-reversal symmetry by non-zero
$\theta$ increases the conductance in the system.  

\begin{figure}[t]
\begin{center}
\includegraphics[width = 6cm]{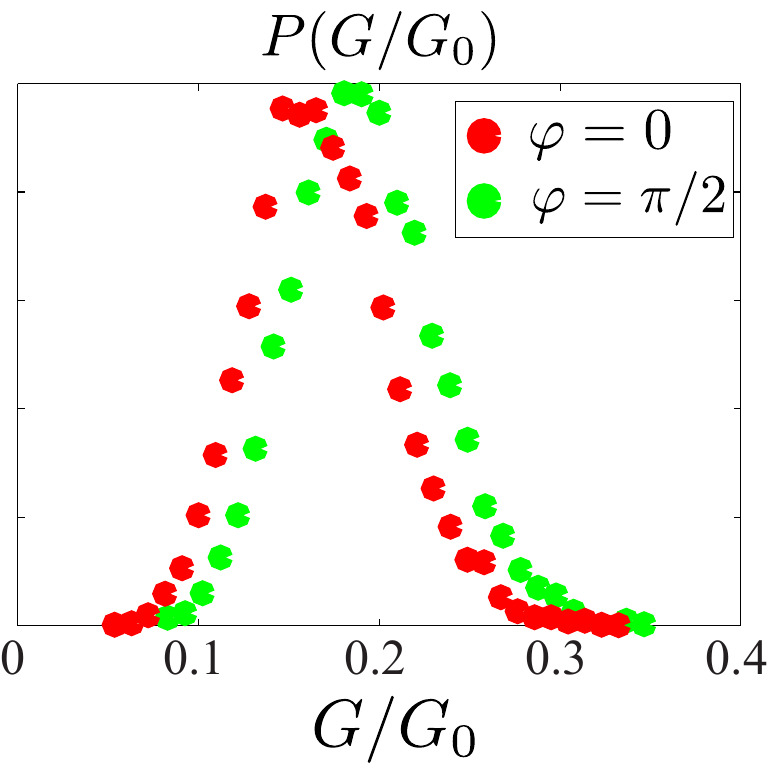}
\caption{ The conductance distributions of disordered systems in the presence of the driving $E \left\{ \cos(\Omega t)  + \cos(2\Omega t + \phi) \right\}$ with $\phi=0$ and $\phi=\pi/2$. 
Here, $L=601$, $W=0.6 J$, $\Omega=0.01 J$ and intensity of light $\mathcal{A} =0.3$. 
While the average value is increased by the breaking of time-reversal symmetry, 
the shape of the distributions remain Gaussian.}
\label{TR_comparison}
\end{center}
\end{figure}

On the other hand, the breaking of time reversal symmetry seems to have little effects on the 
shape of the distribution. 
In \fref{TR_comparison}, we give the conductance distributions for disordered systems 
for the same parameter as in \fref{theta_variation} for $\theta = 0 $ and $\theta = \pi/2$. 
The distribution of the conductance 
 is always Gaussian, irrespective of the value of $\theta$.

In \fref{TR_comparison}, we give the conductance distributions for disordered systems with $\Omega$
and $2\Omega$ drives with phase difference $\phi=0$ and $\phi=\pi/2$. Here we choose $L=601$, 
$W=0.6 J$, $\Omega=0.01 J$ and intensity of light $\mathcal{A} =0.3$. To ensure the convergence of 
the conductance, we take up to $\pm 50$ Floquet levels in the numerical computations. 
Conductance distributions of 
time-reversal symmetric drive and time-reversal symmetry breaking drive 
have almost identical shape, but their average values are different by about $20 \%$, where 
time-reversal symmetry breaking drive leads to lower conductance. 
The difference in the average conductance decreases as the intensity of drive $\mathcal{A}$ 
is decreased. 

While in one dimensional disordered systems we study here, the time-reversal symmetry breaking only led to a small 
change of conductance, the idea is general and can be applied to higher dimensions. For example, 
application of linearly polarized light to two dimensional system does not break time-reversal symmetry,
but that of circularly polarized light breaks the time reversal symmetry\cite{Kitagawa2011}. 
Thus such control of symmetry class of driven systems is a promising approach to control the conductance 
properties.

 \subsection{Conductance dependence of a sample on the driving frequency and intensity: \\
 giant opto-response of conductance. } \label{sec:giant_response}
 \subsubsection{Frequency dependence}
In the previous sections, we analyze the conductance distribution of disordered systems under the
application of light. 
In experiments, it might not be easy to create so many samples with different realization of disorders,
and question arises as to how one can measure such conductance distributions. 
 In the static, weakly disordered systems, one aspect of conductance distribution, namely fluctuation, is 
 probed as a function of magnetic field\cite{Umbach1984, Lee1985}. 
While such study is, in principle, different from the study of fluctuations for different realizations of disorders, 
it probes inherent fluctuation structure built in the eigen-energy distributions of disordered systems. 
As a result, the study of fluctuations for both different realizations of disorders and intensity of magnetic field 
displays universal conductance fluctuations\cite{Lee1985, Altshuler1985}
 
  In disordered systems in dimensions higher than one, 
  the application of magnetic field might be used to study fluctuations. However, 
 in one dimensional system, there is no orbital effect of magnetic field on the conductance.
 If we suppose that the wire is running along $x$ direction and magnetic field is 
 pointing in $z$ direction, the operation $y \rightarrow -y$ reverses the direction of magnetic field,
 but does not change the current for spin-less electrons, and thus we see that magnetic field does not 
 have any effect on conductance. 
 
 Here we propose the study of conductance fluctuations of disordered systems under the application of light
 as a function of driving frequency. We note that such study is essentially {\it different} from the study of 
 conductance fluctuations for different realizations of disorders, but yet probes inherent fluctuation structure 
 in the disordered systems. Apart from the fundamental question of conductance fluctuations, this study is 
 also interesting from application point of view as a knob to change the conductance of disordered systems. 
 
 \begin{figure}[t]
\begin{center}
\includegraphics[width = 8.5cm]{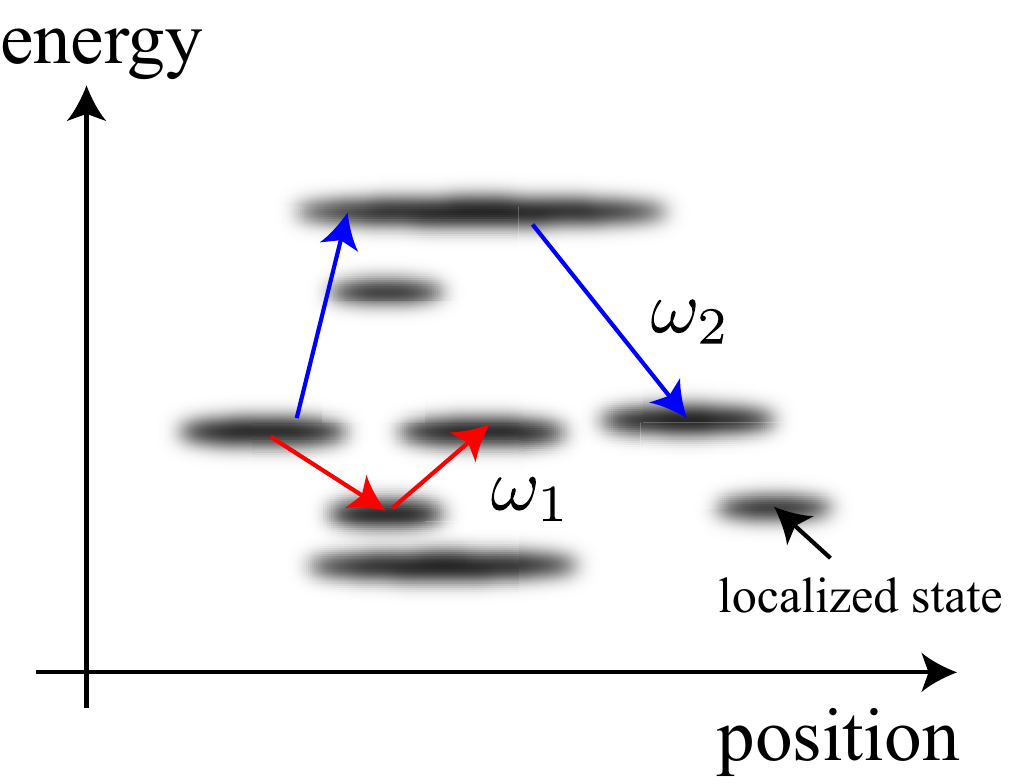}
\caption{Illustration of the photon-assisted hopping in the strongly disordered systems. 
Each state with a given energy $E$ is spatially localized. If two states are spatially overlapping, and moreover
two states are roughly separated by $\Omega$ where $\Omega$ is the 
driving frequency,  the electrons can hop from one state to the other. The energy of each state is widened by the intensity of light $J \mathcal{A} $, where $\mathcal{A}$ is the intensity of light and $J$ is the static hopping amplitude.  
Changing of driving frequencies from, say, $\omega_{1}$ to $\omega_{2}$ in this picture,  leads to
coupling of different states, resulting in the large change of conductance for a given realization of disorders. }
\label{photon_hop}
\end{center}
\end{figure}

In strongly disordered systems under the application of light, localized electrons can hop to neighboring localized states
with the help of photon energy, see \fref{photon_hop}. In this picture, the accessible states are determined by two
conditions; 1. the spatial overlap of 
localized states; 2. energy separation between the two localized states. For the second condition, two states need to be
separated by the frequency of light $\Omega$. This resonant condition is widened due to the intensity of light,
where the widening energy is given by 
Rabi frequency of drive roughly given by $J \mathcal{A}$ for weak intensity of light. 
Thus the electron can hop from one state to the other 
if two states are separated by $\Omega - J \mathcal{A} \sim \Omega + J \mathcal{A}$.
While the change of the frequency of drive by a small amount barely changes the conductance averaged over different 
realizations of disorders, such change can dramatically modify the accessible states for a given realization of disorder. 

\begin{figure}[t]
\begin{center}
\includegraphics[width = 8.5cm]{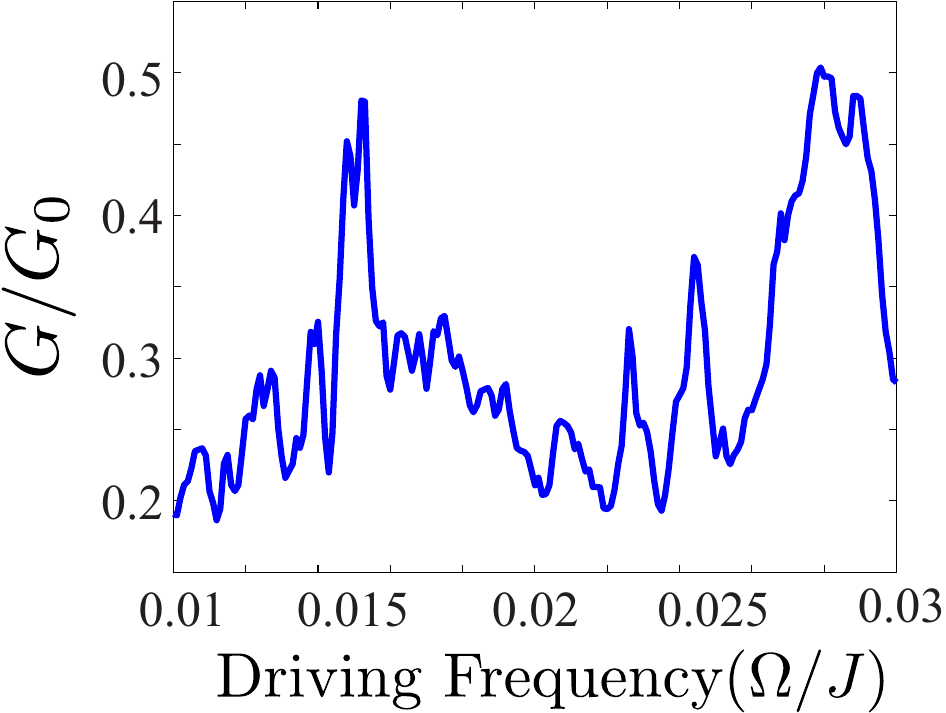}
\caption{ Dependence of conductance on driving frequency 
for the system with a given realization of disorder.  
Here we chose $L=601$ with two channels as before, with intermediate disorder
strength $W=0.6 J$ and small intensity of light $\mathcal{A} = 0.02$.
We observe a giant response of the conductance due to the change of 
frequency by a small amount.}
\label{frequency_dependence}
\end{center}
\end{figure}

In \fref{frequency_dependence}, we plot a dependence of conductance for a given realization of disorder 
for various frequency of drive. Here we chose $L=601$ with two channels as before, with intermediate disorder
strength $W=0.6 J$ and small intensity of light $\mathcal{A} = 0.02$. Despite of small intensity of light, 
the conductance strongly fluctuate as a function of driving frequency $\Omega$. 
The fluctuation is expected to occur at the typical energy spacing of the system, which can be roughly estimated
as $\Delta E = J/L \approx 1.6 \times 10^{-3} J$. 
Such giant response of conductance on the small change of driving frequency could provide an interesting device. 
Note we expect that that the amplitude of fluctuation as a function of driving frequency goes to zero as 
the intensity of light $\mathcal{A}$ goes to zero. Thus for very small intensity of light, the fluctuation amplitude
will be suppressed. 

\subsubsection{ Conductance dependence for varying intensity of light}
In this paper, we study the dependence of conductance distributions on the intensity of light as a general
characterization of statistical ensemble of disordered systems. 
From the point of view of controlling the conductance of disordered systems, yet another interesting 
question is the dependence of a sample with a given disorder on the intensity of light. 
As we have explained above, change of the intensity of light can alter which and how each localized 
states are coupled with each other. This, in return, can change the conductance. 

\begin{figure}[t]
\begin{center}
\includegraphics[width = 8.5cm]{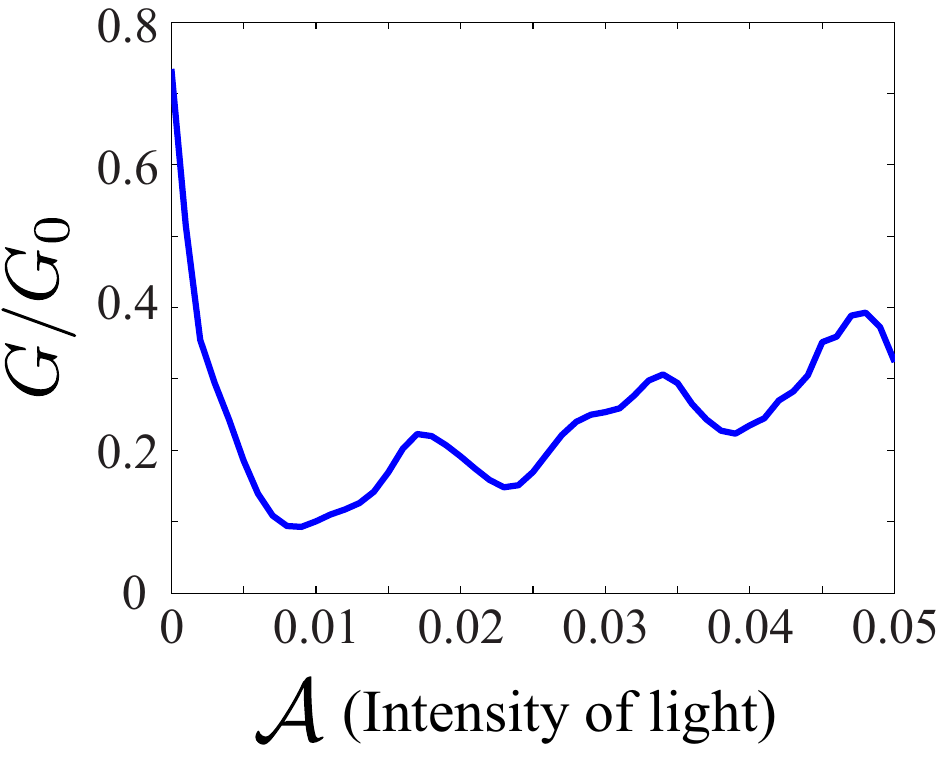}
\caption{Dependence of conductance on the intensity of light, $\mathcal{A}$
for the system with a given realization of disorder. We choose 
$L=601$ with two channels with intermediate disorder
strength $W=0.6 J$ and small driving frequency of light $\Omega = 0.01 J$.
Conductance has a smooth dependence on $\mathcal{A}$, but yet, we observe
a large change of conductance for a small change of $\mathcal{A}$. }
\label{intensity_dependence}
\end{center}
\end{figure}

In \fref{intensity_dependence}, we plot the behavior of conductance for a given realization of disorder 
as a function of the intensity of light, $\mathcal{A}$. We choose 
$L=601$ with two channels with intermediate disorder
strength $W=0.6 J$ and small driving frequency of light $\Omega = 0.01 J$.
Since the intensity of light $\mathcal{A}$ controls the width of resonant condition, the effect is expected to
be more smooth than that of frequency change. Yet, we see in \fref{intensity_dependence} that the 
conductance does change by a large amount for a small change of $\mathcal{A}$. Thus this parameter 
can also be used to induce the giant optical change of conductance for disordered systems. 

\section{Conclusion}
In this paper, we investigated the behavior of full conductance distributions 
of disordered quasi-1D systems under the application of light. By developing the 
Floquet transfer matrix method, we study the effect of application of low frequency light
for large systems, where photon-assisted transport plays an important role in the conductance. 
The comprehensive study of average, fluctuations, skewness and shape of conductance distributions 
showed that the application of light influences the conductance in different ways for weakly, intermediate
and strongly disordered regimes. 
We argued that the conductance is determined by the competition between dynamic localization
which dominates in the weakly localized regime and dimensional increase which dominates in the strongly 
localized regimes. Here we confirmed the previous result\cite{Martinez2006} 
which showed the elongation of localization length
upon the application of light, and saturation of such increase for larger intensity of light in the strongly localized regime.
In the intermediate disorder strength, the application of light results in the crossover behavior from 
log-normal like distribution to normal distribution of conductance. 

Many of the ideas demonstrated in this work can be directly extended to higher dimensional disordered systems. 
The physics of localization is known to strongly depend on the dimensionality of the systems, where 
in 1 and 2D, all the states are localized and in 3D, it is believed that the states are localized below a mobility edge
and extended above it. Thus the effect of dimensional increase is expected to have a large effect in two dimensional 
systems under the application of light. This physics will be studied in the future work. 

In addition, it is interesting to study the disordered two dimensional systems in different 
geometrical structure. For example, hexagonal lattice structure has Dirac cones at the half filling, 
and intriguing weak anti-localization behavior has been observed in graphene\cite{Wu2007}. Linearly polarized
light does not destroy the Dirac cones, and it is of interests to see how such anti-localization is modified 
under the application of light. 

Furthermore, the application of circularly polarized light on hexagonal lattice breaks time-reversal symmetry,
and is predicted to give rise to 
band structures with non-trivial topology which has non-zero Chern numbers\cite{Kitagawa2011}. 
Such non-zero Chern number is known to be the origin of integer quantum Hall effect and chiral edge states. 
Thus in the disordered hexagonal lattice systems under the application of light, one can study the rich physics 
of the interplay between the localization and topological band structures.  

We want to thank a useful discussion with Bertrand Halperin, Jay D Sau, Gil Rafael, Chris R. Laumann, 
and Rarael A. Molina.
We acknowledge the support from Army Research Office with funding from the DARPA OLE
program, Harvard-MIT CUA, NSF Grant No. DMR-07-05472,
AFOSR Quantum Simulation MURI,
the ARO-MURI on Atomtronics.

\appendix

\section{Alternative gauge: \\ sinusoidally varying electric field}
As we have mentioned in the main text, it is sometimes more convenient to use 
the equivalent description of the light application through a different gauge choice. 
If we represent the light field as AC electric field, then the Hamiltonian takes the form
\begin{equation}
H(t) = -J \sum_{<ij>} c_{i}^{\dagger} c_{j}  + \sum_{i} W_{i} c^{\dagger}_{i} c_{i} 
+ E \sum_{i} x_{i} \cos(\Omega t) c^{\dagger}_{i} c_{i} 
\end{equation}
In this gauge, Hamiltonian has only the first harmonics of driving frequency $\Omega$
with $H(1) = H(-1) = \frac{E}{2} \sum_{i} x_{i} c^{\dagger}_{i} c_{i} $. In a finite system, 
this operator is well-defined, but due to the presence of potential proportional to $x_{i}$, 
the interpretation of the operator becomes subtle in the thermodynamic limit. Therefore,
this gauge is computationally useful in a finite system, but the other gauge choice we have 
used in \eqnref{hamiltonian} has clearer conceptual understandings. 

In this gauge, the Schr\"odinger equation takes the form
 \begin{eqnarray*} 
  (E + \hat{\Omega} - \hat{h}_{i}) \vec{s}_{i} &=& - J (\vec{s}_{i-1} +  \vec{s}_{i+1}) \\
   \hat{\Omega} &=& \left(
\begin{array}{ccccc}  \ddots  & &   &  &   \\ &1 \Omega& && 
\\    &  & 0 \Omega & &
 \\  & & & -1 \Omega &   \\  & & &   & \ddots 
\end{array} \right)  \\
\hat{h}_{i} &=&  \left(
\begin{array}{ccccc}  \ddots  & &   &  &   \\ &W_{i} &E x_{i}/2  && 
\\    & E x_{i}/2  & W_{i} & E x_{i}/2 &
 \\  & & E x_{i}/2 & W_{i} &   \\  & & &   & \ddots 
\end{array} \right),
  \end{eqnarray*}
leading to the local transfer matrix 
 \begin{eqnarray*} 
  \left( \begin{array}{c} \vec{s}_{i+1} \\ \vec{s}_{i} \end{array} \right) &=& 
\hat{M}_{i} \left( \begin{array}{c} \vec{s}_{i}  \\ \vec{s}_{i-1}  \end{array} \right) \\
\hat{M}_{i} &=&
\left( \begin{array}{cc} -(E + \hat{\Omega} - \hat{h}_{i})/J  & -1
\\ 1 & 0  \end{array} \right)
\end{eqnarray*}

The transfer matrices that connects the left lead to the central system $\hat{M}_{1}$ and 
the central system to the right lead $\hat{M}_{L}$ are given by 
 \begin{eqnarray*} 
\hat{M}_{1} &=&
\left( \begin{array}{cc} -(E + \hat{\Omega} - \hat{h}_{i})/J  & -J_{L}/J
\\ 1 & 0  \end{array} \right) \\
\hat{M}_{L} &=&
\left( \begin{array}{cc} -(E + \hat{\Omega} - \hat{h}_{i})/J_{L}  & -J/J_{L}
\\ 1 & 0  \end{array} \right) 
\end{eqnarray*}

As in the other gauge, the transfer matrix $\hat{M}$ is now given by 
$\hat{M} = \hat{Q}^{-1} \prod_{i} \hat{M}_{i}  \hat{Q}$ up to the unitary rotation. 
The calculation of the transmission matrix $\hat{t}$ will be analogous as describedi n the main text.


\begin{thebibliography}{10}

\bibitem{Anderson1958}
P.~W. Anderson,
\newblock Phys. Rev. {\bf 109}, 1492 (1958).

\bibitem{Mott}
N.~F. Mott,
\newblock Philosophical Magazine {\bf 19}, 835 (1969).

\bibitem{Ambegaokar1971}
V.~Ambegaokar, B.~I. Halperin, and J.~S. Langer,
\newblock Phys. Rev. B {\bf 4}, 2612 (1971).

\bibitem{Basko2006}
D.~Basko, I.~Aleiner, and B.~Altshuler,
\newblock Annals of Physics {\bf 321}, 1126  (2006).

\bibitem{Wu2007}
X.~Wu, X.~Li, Z.~Song, C.~Berger, and W.~A. de~Heer,
\newblock Phys. Rev. Lett. {\bf 98}, 136801 (2007).

\bibitem{McCann2006}
E.~McCann {\em et~al.},
\newblock Phys. Rev. Lett. {\bf 97}, 146805 (2006).

\bibitem{Wiersma1997}
D.~S. Wiersma, P.~Bartolini, A.~Lagendijk, and R.~Righini,
\newblock Nature {\bf 390}, 671 (1997).

\bibitem{Chabanov2000}
A.~A. Chabanov, M.~Stoytchev, and A.~Z. Genack,
\newblock Nature {\bf 404}, 850 (2000).

\bibitem{Schwartz2007}
T.~Schwartz, G.~Bartal, S.~Fishman, and M.~Segev,
\newblock Nature {\bf 446}, 52 (2007).

\bibitem{Roati2008}
G.~Roati {\em et~al.},
\newblock Nature {\bf 453}, 895 (2008).

\bibitem{Billy2008}
J.~Billy {\em et~al.},
\newblock Nature {\bf 453}, 891 (2008).

\bibitem{Abrahams1979}
E.~Abrahams, P.~W. Anderson, D.~C. Licciardello, and T.~V. Ramakrishnan,
\newblock Phys. Rev. Lett. {\bf 42}, 673 (1979).

\bibitem{Beenakker1997}
C.~W.~J. Beenakker,
\newblock Rev. Mod. Phys. {\bf 69}, 731 (1997).

\bibitem{Martinez2006}
D.~F. Martinez and R.~A. Molina,
\newblock Phys. Rev. B {\bf 73}, 073104 (2006).

\bibitem{Dunlap1986}
D.~H. Dunlap and V.~M. Kenkre,
\newblock Phys. Rev. B {\bf 34}, 3625 (1986).

\bibitem{Grossmann1991}
F.~Grossmann, T.~Dittrich, P.~Jung, and P.~H\"anggi,
\newblock Phys. Rev. Lett. {\bf 67}, 516 (1991).

\bibitem{Keay1995}
B.~J. Keay {\em et~al.},
\newblock Phys. Rev. Lett. {\bf 75}, 4102 (1995).

\bibitem{Lignier2007}
H.~Lignier {\em et~al.},
\newblock Phys. Rev. Lett. {\bf 99}, 220403 (2007).

\bibitem{Holthaus1995}
M.~Holthaus, G.~H. Ristow, and D.~W. Hone,
\newblock Phys. Rev. Lett. {\bf 75}, 3914 (1995).

\bibitem{Holthaus1996}
M.~Holthaus and D.~W. Hone,
\newblock Philosophical Magazine Part B {\bf 74}, 105 (1996).

\bibitem{Lee1985}
P.~A. Lee and A.~D. Stone,
\newblock Phys. Rev. Lett. {\bf 55}, 1622 (1985).

\bibitem{Altshuler1985}
B.~L. Al'tshuler,
\newblock JETP Letters {\bf 41}, 648 (1985).

\bibitem{Gopar2010}
V.~A. Gopar and R.~A. Molina,
\newblock Phys. Rev. B {\bf 81}, 195415 (2010).

\bibitem{Martinez2003}
D.~F. Martinez,
\newblock Journal of Physics A: Mathematical and General {\bf 36}, 9827 (2003).

\bibitem{Kondov2011}
S.~S. Kondov, W.~R. McGehee, J.~J. Zirbel, and B.~DeMarco,
\newblock Science {\bf 334}, 66 (2011).

\bibitem{Chen2011}
Y.-A. Chen {\em et~al.},
\newblock Phys. Rev. Lett. {\bf 107}, 210405 (2011).

\bibitem{Mello2004}
P.~A. Mello and N.~Kumar,
\newblock {\em {Quantum Transport in Mesoscopic Systems: Complexity and
  Statistical Fluctuations}} (Oxford University Press, USA, 2004).

\bibitem{Jauho1994}
A.-P. Jauho, N.~S. Wingreen, and Y.~Meir,
\newblock Phys. Rev. B {\bf 50}, 5528 (1994).

\bibitem{Kohler2005}
S.~Kohler, J.~Lehmann, and P.~HÌ?nggi,
\newblock Physics Reports {\bf 406}, 379  (2005).

\bibitem{Moskalets2002}
M.~Moskalets and M.~B\"uttiker,
\newblock Phys. Rev. B {\bf 66}, 205320 (2002).

\bibitem{Kitagawa2011}
T.~Kitagawa, T.~Oka, A.~Brataas, L.~Fu, and E.~Demler,
\newblock Phys. Rev. B {\bf 84}, 235108 (2011).

\bibitem{Martinez2008}
D.~F. Martinez, R.~A. Molina, and B.~Hu,
\newblock Phys. Rev. B {\bf 78}, 045428 (2008).

\bibitem{Markos2006}
P.~Markos,
\newblock p. 561 (2006).

\bibitem{Kitagawa2010}
T.~Kitagawa, E.~Berg, M.~Rudner, and E.~Demler,
\newblock Phys. Rev. B {\bf 82}, 235114 (2010).

\bibitem{Umbach1984}
C.~P. Umbach, S.~Washburn, R.~B. Laibowitz, and R.~A. Webb,
\newblock Phys. Rev. B {\bf 30}, 4048 (1984).

\end{thebibliography}
\end{document}